\documentstyle[psfig]{mn}
\newcommand{\Msun}{M_{\odot}}
\newcommand{\Zsun}{Z_{\odot}}

\newcommand{\average}[1]{\langle\rm{#1}\rangle_{\rm V}}
\newcommand{\mathaverage}[1]{\langle{#1}\rangle_{\rm V}}
\newcommand{\averagemass}[1]{\langle\rm{#1}\rangle_*}

\newcommand{\etal}{{et~al.}}
\newcommand{\nucl}[2]{\,^{#1}{\rm {#2}}}
\newcommand{\tausf}{\tau_{\rm SF}}
\newcommand{\taupre}{\tau_{\rm pre}}
\newcommand{\taubu}{\tau_{\rm burst}}
\newcommand{\Ae}{$\alpha$-enhancement}

\title[$\alpha$-enhancement in elliptical galaxies]
{\boldmath Constraints on galaxy formation from $\alpha$-enhancement in
luminous elliptical galaxies}

\author[D.~Thomas, L.~Greggio and R.~Bender]
{D.~Thomas,$^{1}$ L.~Greggio$^{1,2}$ and R.~Bender$^{1}$\\
$^1$Universit\"ats-Sternwarte M\"unchen, Scheinerstr.~1, D-81679 M\"unchen,
Germany\\
$^2$Dipartimento di Astronomia, Universit\`{a} di Bologna, I40100 Bologna,
Italy}

\begin{document}

\maketitle
\begin{abstract}
We explore the formation of $\alpha$-enhanced {\em and} metal-rich stellar
populations in the nuclei of luminous ellipticals under the assumption of
two extreme galaxy formation scenarios based on hierarchical clustering,
namely a {\em fast clumpy collapse} and the {\em merger of two spirals}. We
investigate the parameter space of star formation time-scale, IMF slope, and
stellar yields. In particular, the latter add a huge uncertainty in
constraining time-scales and IMF slopes. We find that -- for Thielemann,
Nomoto \& Hashimoto \shortcite{TNH96} nucleosynthesis -- in a {\em fast
clumpy collapse} scenario an [$\alpha$/Fe] overabundance of $\sim 0.2$ dex
in the high metallicity stars can be achieved with a Salpeter IMF and star
formation time-scales of the order $10^9$ yr. The scenario of two {\em
merging spirals} which are similar to our Galaxy, instead, fails to
reproduce $\alpha$-enhanced abundance ratios in the metal-rich stars, unless
the IMF is flattened during the burst ignited by the merger. This result is
independent of the burst time-scale. We suggest that abundance gradients
give hints to distinguish between the two extreme formation scenarios
considered in this paper.
\end{abstract}


\section{Introduction}
\label{intro}
An enhancement of $\alpha$-elements\footnote{The so-called
$\alpha$-elements, i.e.\ O, Mg, Si, Ca, Ti, are build up by synthesizing
$\nucl{4}{He}$-particles.} in the {\em metal-poor} stars of our Galaxy, with
respect to the solar proportions, was first realized by Aller \& Greenstein
\shortcite{AG60}, Wallerstein
\shortcite{Wa62} and Conti \etal\ \shortcite{Cetal67}.
Subsequent studies (see Truran \& Burkert 1993; Gehren 1995, and references
therein)\nocite{TB93,G95} have confirmed that the abundance ratios in
metal-poor stars exhibit $\alpha$/Fe larger than the solar ratios. Since
$\alpha$-elements are mainly produced in short-lived, massive stars, while a
substantial contribution to Fe comes from Type~Ia supernovae (SN~Ia) with
longer lived progenitors, this pattern is generally interpreted in terms of
formation time-scales. In the early stages of the chemical evolution (i.e.\
at low metallicities) the stellar abundance ratios reflect the SN~II
products. As the metallicity in the interstellar medium (ISM) builds up
towards the solar value, the [Mg/Fe] ratio is significantly driven down by
the products of SN~Ia explosions (cf.\ McWilliam 1997 and references
therein\nocite{McW97}). Abundance ratios therefore carry information about
the main contributors to the enrichment and the star formation time-scales
(SFT).

In contrast to the Milky Way, it is generally impossible to resolve stars in
elliptical galaxies, thus their chemical properties have to be inferred from
spectral indices. Theoretical population synthesis models based on {\em
solar} abundance ratios, while adequately describing the $\langle {\rm
Fe}\rangle$-Mg$_2$ features observed in the low luminosity ellipticals, fail
to reproduce these indices in brighter objects (Peletier 1989; Gorgas,
Efstathiou \& Arag{\'{o}}n Salamanca 1990; Faber, Worthey \& Gonz\'{a}lez
1992; Worthey, Faber \& Gonz\'{a}lez 1992; Davies, Sadler \& Peletier 1993).
\nocite{P89,GAS90,FWG92,WFG92,DSP93}
Specifically, at a given $\langle {\rm Fe}\rangle$ index the Mg$_2$ index is
stronger than predicted by such models. This result is usually
interpreted in terms of an [Mg/Fe] overabundance in the stellar
populations inhabiting the most massive ellipticals.  
The following explanations for this $\alpha$-enhancement have been proposed 
(e.g.\ Faber \etal\ 1992)\nocite{FWG92}:
\begin{enumerate}
\item Short time-scales characterizing the process of formation of the 
stellar populations (SP) dominating the observed line indices.
\item Selective mass loss, such that SN~Ia products are lost more efficiently.
\item An Initial Mass Function (IMF) biased towards massive stars.
\item Lower rate of SN~Ia with respect to what is appropriate for the Milky
Way.
\end{enumerate}
In cases (i) and (ii) the intracluster medium (ICM) should be overabundant
in iron, if the contribution of giant ellipticals (hereafter gEs) to the ICM
pollution is large \cite{Retal93}. Recent results from ASCA indicate that
the element ratios in the ICM are consistent with solar values (Mushotzky
\etal\ 1996; Ishimaru \& Arimoto 1997; Gibson, Loewenstein \& Mushotzky 1997;
Renzini 1997)\nocite{Metal96,IA97,GLM97,R97}. This suggests that either gEs
do not play a major role in establishing the ICM composition or options (i)
and (ii) do not apply.
Case~(iv) is an ad hoc assumption without real justification. 
Observed nova rates and LMXB populations
make it unlikely that early-type galaxies form less binary stars than later 
types \cite{Da96}. 

Options (i) and (iii) have been investigated numerically by several authors
\cite{M94,Tetal96,Vetal96,Vetal97,GM97,Cetal97} in the
framework of monolithic collapse models for the galaxy formation. 
These models describe the elliptical galaxy as a single body, and
their results are hence appropriate to describe the {\em global} properties of
these objects. These studies concentrate on reproducing the trend of
$\alpha$-enhancement increasing with galactic mass, that can be inferred
from the observed spectral indices. The main results of these
computations are:
\begin{enumerate}
\item The classical wind models yield {\em decreasing} [Mg/Fe] overabundance
with increasing galactic mass in contradiction to the observations.
\item An increasing [Mg/Fe] with mass can be reproduced if
the SFT becomes shorter with increasing mass of the elliptical. This is the
case for the inverse wind models proposed by Matteucci \shortcite{M94}.
\item Alternatively, an IMF getting flatter with increasing galactic mass
reproduces the same pattern. 
\end{enumerate}
The most recent works on this subject tend to favour the flat IMF option for
gEs \cite{Vetal97,GM97,Cetal97}.

\smallskip
In this paper we re-investigate options (i) and (iii). Rather than
addressing the question of the increasing overabundance with galactic mass,
we seek under which conditions an overabundance at high $Z$ -- appropriate
for nuclei of gEs -- is produced. In doing so we consider a physically
plausible scenario for the formation of giant ellipticals based on
hierarchical clustering of CDM theory \cite{WR78,Fetal85,Eetal88}.

Indeed there are several indications that merging must play an important
role in galaxy formation. Approximately 50 per cent of luminous ellipticals
host kinematically decoupled cores \cite{Be90,Be96}. In some cases, the SPs
in these cores show markedly different properties in the spectral indices
\cite{BS92,DSP93}. As shown by Greggio (1997, hereafter G97)\nocite{Gre97},
the strong Mg$_2$ and $\langle {\rm Fe}\rangle$ indices measured in the
nuclei of gEs indicate a substantial degree of pre-enrichment in the gas
which is converted into stars in the inner parts of these galaxies. Although
not exclusively, this can be accomplished in a merging formation scenario,
provided that gas dissipation plays an important role.

Kinematical properties and line
strength features are then determined by the ratio of star formation
and merging
time-scales (Bender \& Surma 1992; Bender, Burstein \& Faber 1992)
\nocite{BS92,BBF92}. In this framework we can envisage two extreme cases for the
formation scenario for ellipticals, mainly
distinguished by the overall formation time-scale of the object:
\begin{description} 
\item[a)] {\em Fast Clumpy Collapse}.
On rather short time-scales ($\sim 1$ Gyr), massive objects are built up by
merging of smaller entities. Star formation (SF) occurs within these
entities as they merge. The different dissipative properties of the newly
formed stars and the gas produce a chemical separation, such that the
enriched material flows to the centre (Bender \& Surma 1992;
G97)\nocite{BS92,Gre97}. On a larger scale, however, the whole system
participates in a general collapse, merging and SF occuring within the same
(short) time-scale. We notice that, from a chemical evolution point of view,
this formation mode can also be considered as a refinement of the classical
Larson \shortcite{L74} monolithic collapse.
\item[b)] {\em Merging Spirals}. The merging event occurs when the
merging entities have already converted most of their gas into stars.
For example, two spirals similar the Milky Way merge to form an elliptical 
galaxy \cite{FS82,NW83,G83,B88,H93}. Most of the stars in the merging
entities have formed in a continuous and long lasting ($\sim$ 10 Gyr)
SF process, leading to approximately solar abundances in the ISM. At
merging, the (enriched) residual gas flows down to the centre \cite{HB91,BH96} 
where it experiences a violent SF episode, with a short time-scale. The
global formation process, however, lasts $\sim$ 10 Gyr or more.
\end{description}
Semi-analytic models of galaxy formation in the framework of the CDM
theory for structure formation (Kauffmann, White \& Guiderdoni 1993; Lacey
\etal\ 1993; Cole \etal\ 1994)\nocite{KWG93,Letal93,Cetal94}
provide a continuum between these two extremes, with a distribution 
of time-scales for the formation of ellipticals which depends on specific 
assumptions in the modeling.
There are several arguments constraining short formation time-scales 
and old formation ages of elliptical galaxies, hence pointing towards
the {\em fast clumpy collapse} model.
They are mainly based on the tightness of 
Fundamental Plane relations, i.e.\ the relation between SPs and $\sigma$
\cite{Dretal87,DD87} and the small scatter in $M/L$ ratios perpendicular to the
FP \cite{RC93}. Also the colour evolution \cite{Aetal93}, and the evolution
of the Mg-$\sigma$ relation with redshift (Bender, Ziegler \& Bruzual 1996;
Ziegler \& Bender 1997)\nocite{BZB96,ZB97} put the star formation ages of
{\em cluster} ellipticals to high redshift ($z\ga 2$). Despite this, the
chemical outcome of scenario~(b) is of interest, since we know that such
mergers do occur (IRAS galaxies, e.g.\ Soifer \etal\ 1984; Joseph \& Wright
1985; Sanders \etal\ 1988; Melnick \& Mirabel
1990).\nocite{Setal84,JW85,Setal88,MM90} If ellipticals in the field and in
clusters are intrinsically different (e.g.\ Kauffmann, Charlot \& White
1996)\nocite{KCW96}, and those in clusters form earlier, and on shorter
time-scales \cite{K96}, the two distinct scenarios may be assigned to these
two kinds of objects. We notice, however, that the \Ae\ does not seem to
depend on cluster properties (J{\o}rgensen, Franx \& Kj{\ae}rgaard
1995)\nocite{JFK95}, at least within the errors. Also, the Mg-$\sigma$
relation measured in many clusters is remarkably independent of the clusters
properties \cite{Cetal98}. Finally, the Mg-$\sigma$ relation for field
ellipticals does not differ appreciably from that derived for the Coma
Cluster (Bernardi \etal, in preparation). Despite this, the observational
situation is still far from being settled, and we investigate on the
predictions of both models to help in constraining the problem of the
formation of ellipticals.

The paper is organized as follows.
In Section~\ref{model} we provide a more detailed prescription of the 
theoretical model used in this paper. The results are presented and discussed 
in Section~\ref{results} and Section~\ref{discussion}, respectively.
The main conclusions are summarized in Section~\ref{conclusion}.


\section{The theoretical model}
\label{model}

The basic constraints on our modeling, aimed to describe the SPs in
the central parts of bright ellipticals, are:
\begin{enumerate}
\item Achieving high (super-solar) total metallicities ($Z$);
\item Producing an \Ae\ of the order of
[$\alpha$/Fe]$\sim$ 0.2--0.4 dex, as implied by the observations.
\end{enumerate}
The first requirement comes from the observed nuclear Mg$_2$ index and
points towards either higher yields (i.e.\ shallow IMF) in the central
regions or to a scenario of {\em enriched} inflow (Edmunds 1990, 1992;
G97)\nocite{Ed90,Ed92}. The latter implies that the SPs inhabiting the
nuclear regions of ellipticals are characterized by a metallicity
distribution with a minimum metallicity $Z_{\rm m}>0$ (i.e. they are
composite stellar populations, hereinafter CSP). The second requirement,
instead, comes from both Mg and Fe indices, thus in principle the two
constraints might refer to different components of the CSP. However, since
the Mg and Fe indices are measured in the same spectral range, the component
of the CSP dominating the Mg$_2$ index also dominates the Fe index. {\bf
Hence, requirements (i) and (ii) must be met simultaneously from the SP
dominating the visual light}.

The first constraint is accomplished with a high degree of chemical
processing, hence a large fractional amount of gas has to be turned into
stellar mass. This can be achieved either in short time-scales (perhaps like
in galactic bulges) or long time-scales (as in galactic disks). The second
characteristic points towards relatively short time-scales. Note that both
constraints are generally more easily met when assuming a flatter IMF.

\begin{figure}
\psfig{figure=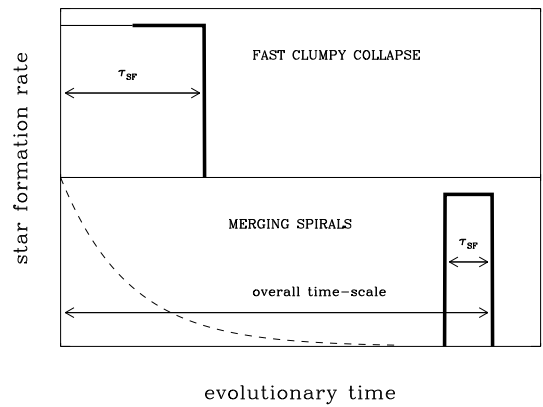,width=\linewidth}
\caption{A sketch of the star formation histories assumed for the {\em fast
clumpy collapse} (top panel) and the {\em merging spirals} (bottom panel)
schemes. The arrows indicate the duration of the SF episode in the modeling
($\tausf$). The thick solid lines indicate the SF episode yielding the SPs
which end up in the central parts of the modeled galaxy. The dashed line
(bottom panel only) sketches the rough star formation history of the parent
spirals before the merger.}
\label{fig:sfh}
\end{figure}

The {\em fast clumpy collapse} and the {\em merging spirals} modes differ
mainly in their time-scales of the pre-enrichment processes. These are
mimicked by assuming different initial conditions for the chemical
processing of the gas which forms the CSP hosted in the central parts of the
galaxy. In the {\em fast clumpy collapse} the process of formation is so
fast that the galaxy can be viewed as a single body. Its chemical evolution
can be described as a closed box with only radial separation of different
SPs. The initial abundance for the SF process is assumed to be primordial.
In the {\em merging spirals} model, instead, the gas which is turned into
stars at the merging event has been enriched in the extended evolution
history of the parent spirals leading to approximately solar abundances. The
two different star formation histories are sketched in Fig.~\ref{fig:sfh}.
The duration of the active SF process will be addressed to as the star
formation time-scale $\tausf$ (arrows in the figure). In the case of the
{\it merging spirals} model (bottom panel) the overall formation time-scale
for the final elliptical is much longer than $\tausf$. The phase of chemical
enrichment in the parent spirals {\em before} the merger is indicated by the
dashed line. The chemical outcome of this phase is reflected in the initial
abundances assumed for the gas participating the merger (see
Table~\ref{tab:pres}). The thick lines in Fig.~\ref{fig:sfh} indicate the
metal-rich SPs that are assumed to form the nucleus of the object. 

In the simulations, we consider a normalized mass of gas and calculate its
chemical evolution until the gas is completely used up, yielding the maximum
metallicity, which is appropriate for the most massive ellipticals. It
should be noticed that in the models by Arimoto \& Yoshii \shortcite{AY87}
and Matteucci \& Tornamb\`{e} \shortcite{MT87} at the onset of the galactic
wind the gas fraction is $0.01-0.05$ in the most massive models, depending on
the IMF slope.

One further distinction between the two extreme cases for elliptical galaxy
formation concerns the contribution of SNe~Ia to the Fe enrichment of the
high $Z$ stars. In the merger of the two spirals we neglect the contribution
from those SNe~Ia whose progenitors were born in the parent spirals (dashed
line in Fig.~\ref{fig:sfh}), and which explode after the merging. In other
words, we assume that the products of these SNe~Ia are not efficiently mixed
down to the central part where the final burst occurs. This corresponds to
maximizing the [Mg/Fe] overabundance achieved in the nuclear CSP. In a {\em
fast clumpy collapse}, instead, the merging entities are in close
interaction, chemical mixing
is likely to be more efficient. The model corresponds to the formation of
the whole elliptical in a single event, and it seems quite artificial to
exclude a fraction of the SN~Ia products from the enrichment.

\subsection{Input ingredients for the chemical evolution}
We adopt a constant SFR in our computations, since in both models the SF
episode generating the high $Z$ stars has to be short to accomplish
significant [Mg/Fe] overabundances. 

The chemical evolution is calculated by solving the usual set of
differential equations as described in Thomas, Greggio \& Bender (1998,
hereafter Paper~I)\nocite{TGB98a}. Enrichment due to planetary
nebulae from low-mass stars \cite{RV81}, SNe~II from high-mass stars
(Woosley \& Weaver 1995, hereafter WW95, and Thielemann, Nomoto \& Hashimoto
1996, hereafter TNH96)\nocite{WW95,TNH96} and SNe~Ia occuring in close
binary systems (Nomoto, Thielemann \& Yokoi 1984)\nocite{NTY84} are taken
into account. The fraction $A$ of close binaries is calibrated on a
simultaneous fit of the age-metallicity relation in the solar neighbourhood
and the ratio $N_{\rm Ia}/N_{\rm II}\approx 5$ typically observed in Sbc
galaxies
\cite{vBT91}, leading to $A=0.035$ (see Paper~I).
The IMF is assumed as the usual single slope power-law normalized by mass,
hence the slope $x=1.35$ refers to the Salpeter value \cite{S55}.
The upper and lower mass cut-offs are $40~\Msun$ and $0.1~\Msun$, respectively.
The effect of adopting a different contribution from SN~Ia, different 
mass cut-offs, and an IMF deviating from the single power law (e.g.\ Scalo
1986)\nocite{S86} are discussed below.

The time-scale $\tausf$, and the IMF slope $x$ represent the main parameters
in the calculations.

\subsection{Average abundances}
\label{model_average}
The output of our models are abundances. For a comparison to the
observational data the luminosity weighted average abundances have to be
derived. The real abundances averaged by mass are larger, since at
constant age metal-poor stars are brighter (e.g.\ G97). We shall compute both
averages to figure out the effect quantitatively.

The mass averaged abundance of element $i$ in a CSP of age $t_g$, 
including living 
stars and remnants, is given by the following eqn.:
\begin{equation}
{\langle X_i\rangle}_{\rm *}(t_g)\; =
\frac{\int_{t_{\rm 0}}^{t_{\rm 1}} X_i(t)\,\psi(t)\; m_{\rm *}^{\rm
SSP}(t_g-t)\: dt}{\int_{t_{\rm 0}}^{t_{\rm 1}} \psi(t)\; 
m_{\rm *}^{\rm SSP}(t_g-t)\: dt}\ ,
\label{eq:average_mass}
\end{equation}
where $t_0$ and $t_1$ are respectively the epochs of start and end of
the active SF process. The quantity $m_{\rm *}^{\rm SSP}(t_g-t)$
is the fraction of mass in stars (living and remnants) that, being
born in $t$, are present at the epoch $t_g$ (SSP means {\em Simple Stellar
Population} of a fixed single age and single metallicity):

\[ m_{\rm *}^{\rm SSP}(t_g-t)= \]
\begin{equation}
\int_{m_{\rm min}}^{m_{\rm to}(t_g-t)} \phi(m)\: dm
+
\int_{m_{\rm to}(t_g-t)}^{m_{\rm max}} w_m\,\phi(m)\: dm
\label{eq:stellar_mass}
\end{equation}
In Eqn.~\ref{eq:stellar_mass} 
$m_{\rm to}$ is the turn-off mass at epoch $t_g$ of the SSP born
at time $t$, $m_{\rm min}$ and $m_{\rm max}$ are respectively the
lower and upper mass cut-offs of the IMF, and $w_{m}$ is the
remnant mass of a star of initial mass $m$. The values of $w_{m}$ as a
function of mass are taken from the nucleosynthesis prescriptions (see
Paper~I).
As usual, the IMF $\phi(m)$ is normalized to 1. 
Thus the product $\psi(t)\times m_{\rm *}^{\rm SSP}(t_g-t)$ gives the mass
contributed at the time $t_g$ from the stellar generation born at
epoch $t$, when the abundance in the ISM was $X_i(t)$.

In the light averaged abundance, the weights are given by the luminosity
contributed at epoch $t_g$ from the SSP born at time $t$:
\begin{equation}
{\langle X_i\rangle}_{\rm V}(t_g)\; =
\frac{\int_{t_{\rm 0}}^{t_{\rm 1}} X_i(t)\psi(t)\; L_{\rm V}^{\rm
SSP}(t_g-t)\: dt}{\int_{t_{\rm 0}}^{t_{\rm 1}} \psi(t)\; L_{\rm V}^{\rm
SSP}(t_g-t)\: dt}
\label{eq:average_light}
\end{equation}
where we have specified the V-band wavelength range because this is
most relevant to our application. 
$L_{\rm V}^{\rm SSP}(t_g-t)$ is the luminosity at epoch $t_g$ of an SSP
of unitary mass with an age of $(t_g-t)$, and is given by SSP
photometric models as a function of age and metallicity. In our
application we will use the models by Worthey \shortcite{Wo94}.
The spread in age and metallicity of the CSP
are $[t_0,t_1]$ and $[Z_{\rm m}(t_0),Z_{\rm\-M}(t_1)]$, 
respectively.

The averaged abundance ratio is obtained by taking the ratio of the
averaged abundances. For the sake of clarity, however, we use the following 
notation:
\begin{equation}
\langle [X_i/X_j]\rangle\equiv [\langle X_i\rangle/\langle X_j\rangle]
\end{equation}
where the square brackets denote the usual normalization to the solar
abundances, for which we use the meteoritic abundances given by Anders
\& Grevesse \shortcite{AG89}.


\section{Results}
\label{results}

In this section we present the results of our computations for the two
extreme models of elliptical galaxy formation.
In particular, we consider the average abundances 15~Gyr after the first
episode of SF. However, since the passive evolution of SPs is most rapid in
the first Gyr, the result is insensitive to the details of the star
formation history and the exact age \cite{Ed92}.

Since most of the data in the literature concern magnesium and iron indices,
we will concentrate on the average abundances of these two elements,
thus on the possibility of producing an [Mg/Fe] overabundance. 
Mg can be considered as a representative for $\alpha$-elements. Indeed, the
ratios [Mg/Fe] and [O/Fe] do not differ significantly in the TNH96 models
(Paper~I). Also, we consider V-luminosity averaged quantities, since
both the Mg and Fe indices fall in this spectral range.

In the following, we first discuss the average abundances for the closed box
model. This is a good description of the {\em global} properties of an
elliptical galaxy formed in a {\it fast clumpy collapse} mode. Second we
look at the abundances of the high metallicity component of the closed box
model. This is assumed to describe the properties of the {\em nuclear regions}
of an elliptical galaxy formed in a {\it fast clumpy collapse} mode. Third
we consider the {\em merging spirals} model in two possible initial
conditions (see Table~\ref{tab:pres}), corresponding to different degrees of
enrichment of the ISM at the merging event, hence to different
pre-enrichment time-scales.
\begin{table}
\caption{The initial abundances of the gas assumed to form the central
stellar populations of the elliptical according to the merger of two
spirals. The values for $0.5~\Zsun$ are adopted from Paper~I. The quantity
$\taupre$ denotes the evolutionary time-scale of the parent spirals before
the merger.}
\begin{tabular}{rrrr}
\hline
$Z_{\rm in}$ & $\taupre$  & [Fe/H]$_{\rm in}$ & [Mg/Fe]$_{\rm in}$ \\
$0.5~\Zsun$   & 3 Gyr  & -0.39             & 0.08 \\
$1.0~\Zsun$      & 10 Gyr &  0.00             & 0.00 \\
\hline
\end{tabular}
\label{tab:pres}
\end{table}

\smallskip
The complete set of numbers for different parameter pairs of $\tausf$, $x$,
and $Z_{\rm m} $ and for both formation scenarios is presented in
Table~\ref{tab:all}.
\begin{table*}
\begin{minipage}{15.6cm}
\caption{The respective average stellar abundances (V-luminosity weighted)
resulting from the {\em fast clumpy collapse} model (columns $4-7$) and the
{\em merging spirals} scheme (columns $8-10$) for the different parameters
IMF slope $x$, star formation time-scale, and minimum metallicity
$Z_{\rm\-m}$. TNH96 yields are adopted. In column~4, the quantity
$m_{\rm\-CSP}$ denotes the mass fraction of the segregated composite stellar
population for the {\em fast clumpy collapse}. In the {\em merging spiral}
model, these fractions are determined by the fractions of gaseous mass when
the merger occurs. The model for the evolution of the solar neighbourhood
(Paper~I) yields 0.5 and 0.2 for $Z_{\rm\-m}=0.5~\Zsun$ and
$Z_{\rm\-m}=1~\Zsun$, respectively.}
\begin{tabular}{|ccc|ccr@{0.}lr@{0.}l|cr@{0.}lc|}
\hline
 & & & \multicolumn{6}{c}{FAST CLUMPY COLLAPSE} &
\multicolumn{4}{c}{MERGING SPIRALS}\\
$x$ & SFT (Gyr) & $Z_{\rm m}~(\Zsun)$ & $m_{\rm CSP}$ & 
$\average{[Mg/Fe]}$ &
\multicolumn{2}{c}{$\average{[Fe/H]}$} & 
\multicolumn{2}{c}{$\mathaverage{[Z]}$} & 
$\average{[Mg/Fe]}$ &
\multicolumn{2}{c}{$\average{[Fe/H]}$} & $\mathaverage{[Z]}$ \\\\
1.35 & 0.3 & 0.0 & 1.0 & 0.27 &$-$& 52 & $-$& 27 & --     & 
\multicolumn{2}{c}{--} & -- \\
     &     & 0.5 & 0.5 & 0.26 &$-$& 20 &    & 03 & $0.18$ & $-$& 12 & $0.05$ \\
     &     & 1.0 & 0.2 & 0.25 &$-$& 03 &    & 19 & $0.09$ &    & 14 & $0.20$ \\
     & 1.0 & 0.0 & 1.0 & 0.20 &$-$& 42 & $-$& 22 & --     &
     \multicolumn{2}{c}{--} & -- \\
     &     & 0.5 & 0.5 & 0.19 &$-$& 12 &    & 06 & $0.15$ & $-$& 07 & $0.07$ \\
     &     & 1.0 & 0.3 & 0.18 &   & 03 &    & 20 & $0.07$ &    & 18 & $0.23$ \\
     & 4.0 & 0.0 & 1.0 & 0.09 &$-$& 24 & $-$& 14 & --     &
     \multicolumn{2}{c}{--} & -- \\
     &     & 0.5 & 0.6 & 0.07 &$-$& 01 &    & 09 & $0.08$ &    & 04 & $0.12$ \\
     &     & 1.0 & 0.3 & 0.06 &   & 17 &    & 22 & $0.04$ &    & 23 & $0.29$ \\
1.00 & 0.3 & 0.0 & 1.0 & 0.34 &$-$& 05 &    & 24 & --     &
\multicolumn{2}{c}{--} & -- \\
     &     & 0.5 & 0.8 & 0.34 &   & 11 &    & 38 & $0.28$ &    & 17 & $0.39$ \\
     &     & 1.0 & 0.7 & 0.33 &   & 20 &    & 47 & $0.21$ &    & 34 & $0.48$ \\
     & 1.0 & 0.0 & 1.0 & 0.28 &   & 07 &    & 31 & --     &
     \multicolumn{2}{c}{--} & -- \\
     &     & 0.5 & 0.8 & 0.28 &   & 20 &    & 43 & $0.24$ &    & 25 & $0.44$ \\
     &     & 1.0 & 0.7 & 0.28 &   & 29 &    & 50 & $0.19$ &    & 41 & $0.53$ \\
     & 4.0 & 0.0 & 1.0 & 0.20 &   & 23 &    & 39 & --     &
     \multicolumn{2}{c}{--} & -- \\
     &     & 0.5 & 0.8 & 0.20 &   & 34 &    & 48 & $0.18$ &    & 38 & $0.50$ \\
     &     & 1.0 & 0.7 & 0.20 &   & 42 &    & 55 & $0.15$ &    & 55 & $0.61$ \\
0.70 & 0.3 & 0.0 & 1.0 & 0.40 &   & 31 &    & 62 & --     &
\multicolumn{2}{c}{--} & -- \\
     &     & 0.5 & 0.9 & 0.40 &   & 40 &    & 70 & $0.36$ &    & 44 & $0.70$ \\
     &     & 1.0 & 0.8 & 0.40 &   & 45 &    & 74 & $0.32$ &    & 55 & $0.75$ \\
     & 1.0 & 0.0 & 1.0 & 0.35 &   & 45 &    & 70 & --     &
     \multicolumn{2}{c}{--} & -- \\
     &     & 0.5 & 0.9 & 0.35 &   & 52 &    & 76 & $0.33$ &    & 55 & $0.76$ \\
     &     & 1.0 & 0.9 & 0.35 &   & 56 &    & 79 & $0.29$ &    & 65 & $0.81$ \\
     & 4.0 & 0.0 & 1.0 & 0.29 &   & 62 &    & 79 & --     &
     \multicolumn{2}{c}{--} & -- \\
     &     & 0.5 & 0.9 & 0.29 &   & 67 &    & 83 & $0.27$ &    & 70 & $0.84$ \\
     &     & 1.0 & 0.9 & 0.29 &   & 71 &    & 86 & $0.25$ &    & 84 & $0.92$ \\
\hline   
\end{tabular}
\label{tab:all}
\end{minipage}
\end{table*}

\subsection{\boldmath Predictions from the closed box model}
\label{results_deg}
As mentioned in the Introduction, an [Mg/Fe] overabundance can be realized
either with short SFTs or with a flat IMF. Correspondingly, different IMF
slopes constrain differently the $\tausf$ parameter. We investigate
quantitatively this $\tausf$-$x$ degeneracy, and consider two prescriptions
for the SN~II nucleosynthesis (WW95 and TNH96). In case of WW95 we adopt the
results from model B which assumes an enhancement of the explosion energy in
high mass stars by a factor 1.5 and therefore compares best to the TNH96
models. It should be noted that model B produces the highest Mg/Fe ratio
among the WW95 calculations (models A,B,C). For a detailed discussion we
refer the reader to Paper~I.

\subsubsection{WW95 yields}
\begin{figure}
\psfig{figure=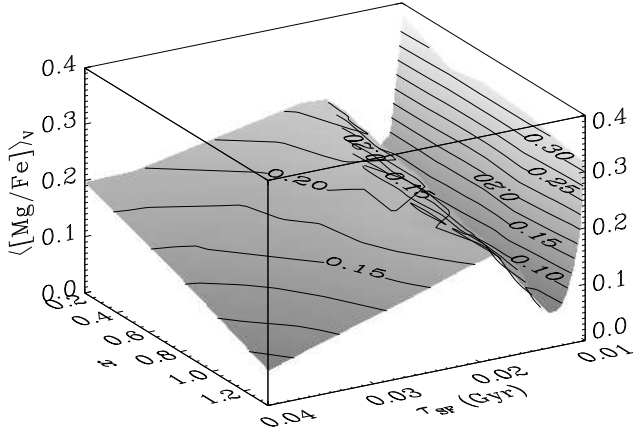,width=\linewidth}
\caption
{The figure shows the average [Mg/Fe] overabundance in the stars weighted by
V-band luminosity as a function of star formation time-scale and IMF slope
(by mass).
SN~II nucleosynthesis is adopted from WW95. The contour lines mark the
respective levels of constant $\average{[Mg/Fe]}$. No segregation of metal-rich
populations is considered, therefore the results are interpreted as {\em
global} properties of the galaxy. The time-scales required for a significant
\Ae\ are unplausible short ($\sim 10^7$ yr).}
\label{fig:ww_surf}
\end{figure}

Fig.~\ref{fig:ww_surf} shows the averaged $\alpha$-enhancement as a function
of SFT and IMF slope for the WW95 stellar yields. The parameter range for
$\tausf$ and $x$ are chosen such that the resulting abundance ratios fall in
the range indicated by the data ($0.2-0.4$ dex). As shown in Paper~I, for
these stellar yields the [Mg/Fe] ratio in the SN~II ejecta from one SSP is
only slightly above solar. It follows that extremely short $\tausf$ are
required to obtain values significantly above solar. In fact,
Fig.~\ref{fig:ww_surf} shows that a $\tausf\sim 10^7$ yr is necessary,
irrespective of the IMF slope. These results are presumably not compatible
with the time-scale to cool the ejecta and to include them in the next
generation of stars.

The minimum in $\average{[Mg/Fe]}$ at $\tausf=11-17$ Myr is a consequence
of the metallicity dependence of the WW95 yields (included in
our computations), in which 
the highest [Mg/Fe] ratios are achieved for lowest ($10^{-4}~\Zsun$) and
highest ($1~\Zsun$) initial metallicity (see tables B1--B12 in Paper~I).
Thus in a regime where SPs of intermediate $Z$ dominate the
enrichment, the calculated overabundance reaches a minimum. 

It should be noticed that the model for the SN~Ia rate adopted in the
computations does not influence this result. The Fe contribution from
these objects starts to play a role only at ages larger than $10^8$ yr.
Lower Fe yields from SN~II would lead to larger $\average{[Mg/Fe]}$ for
a given $\tausf$. However, most of the Fe comes from stars of mass 
$\la 20~\Msun$ for which a lower Fe contribution would be unlikely (see
Paper~I).

In the model by Matteucci \shortcite{M94} an $\alpha$-element overabundance
is still produced with SFTs of the order $10^8$ yr. It is possible that the
different results come from Matteucci \shortcite{M94} having adopted Woosley
\shortcite{W86} yields which do not account for the fall back in high mass
stars, and are therefore characterized by a huge magnesium yield from
massive stars. We have shown in Paper~I (Table~6) that an extrapolation of
the WW95 yields to the high mass end does not increase the Mg abundance in
the ejecta of one SP significantly, again because of the fall back effect.

\subsubsection{TNH96 yields}
\begin{figure*}
\begin{minipage}{0.49\linewidth}
\psfig{figure=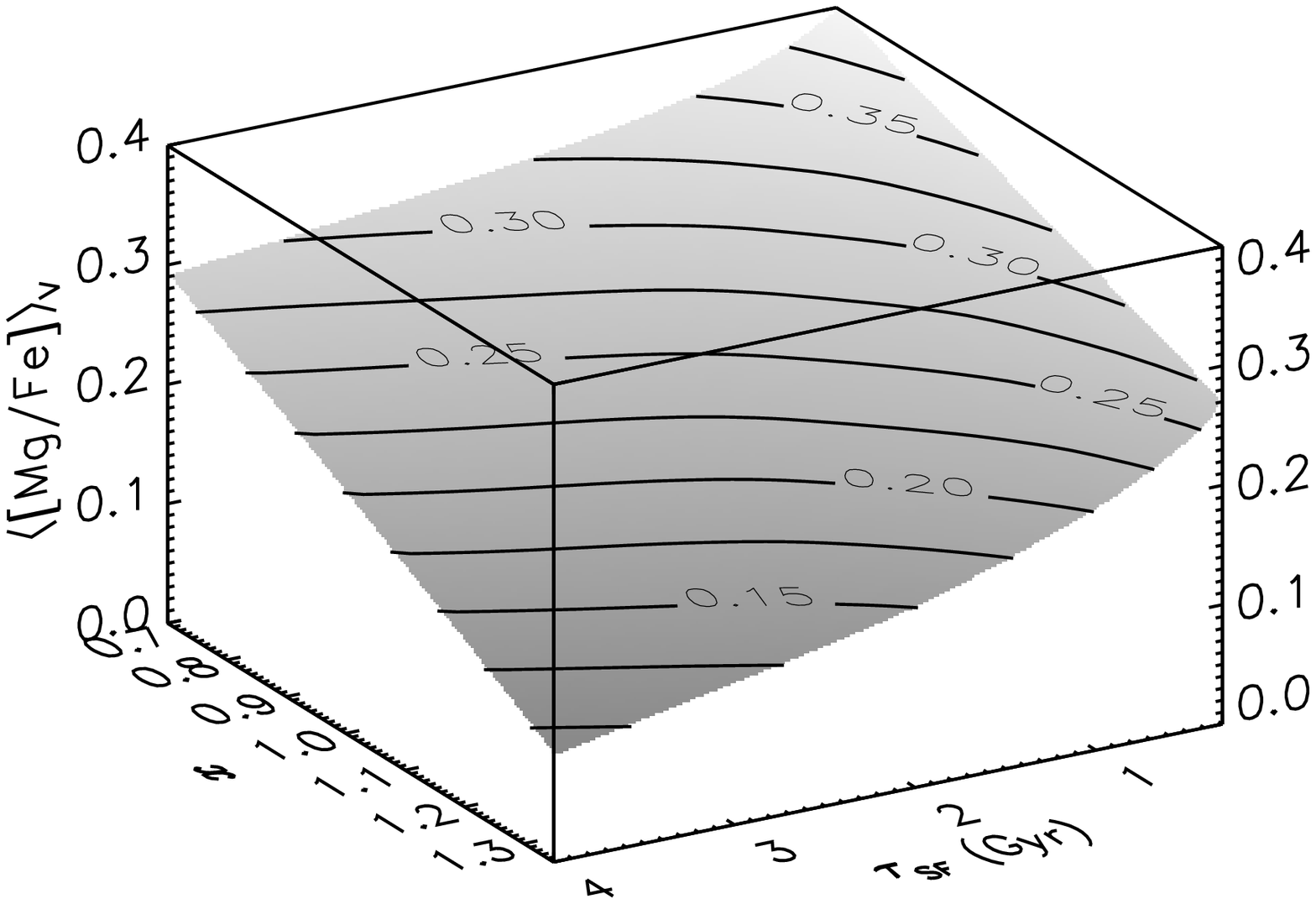,width=\linewidth}
\end{minipage}
\begin{minipage}{0.49\linewidth}
\psfig{figure=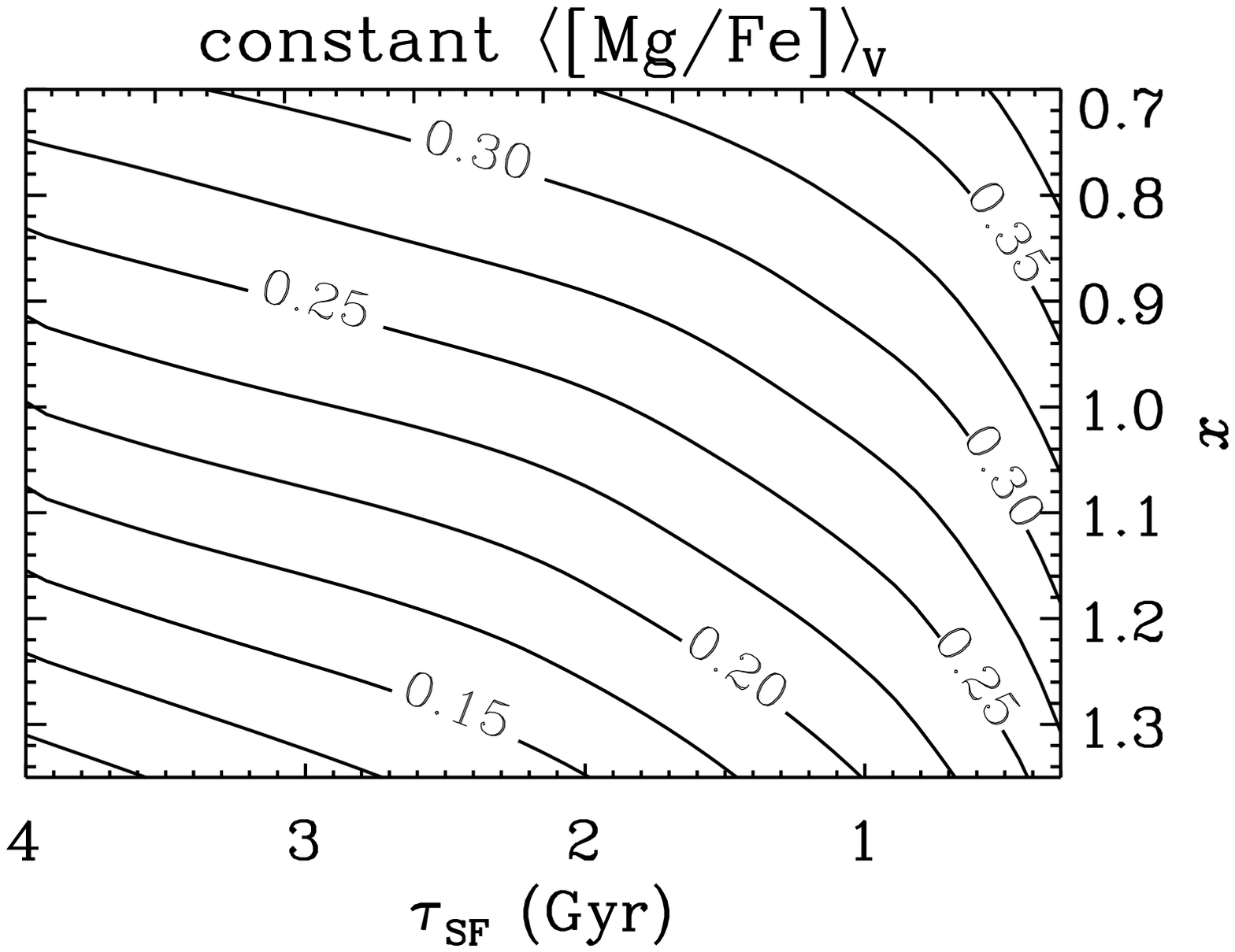,width=\linewidth}
\end{minipage}
\begin{minipage}{0.49\linewidth}
\psfig{figure=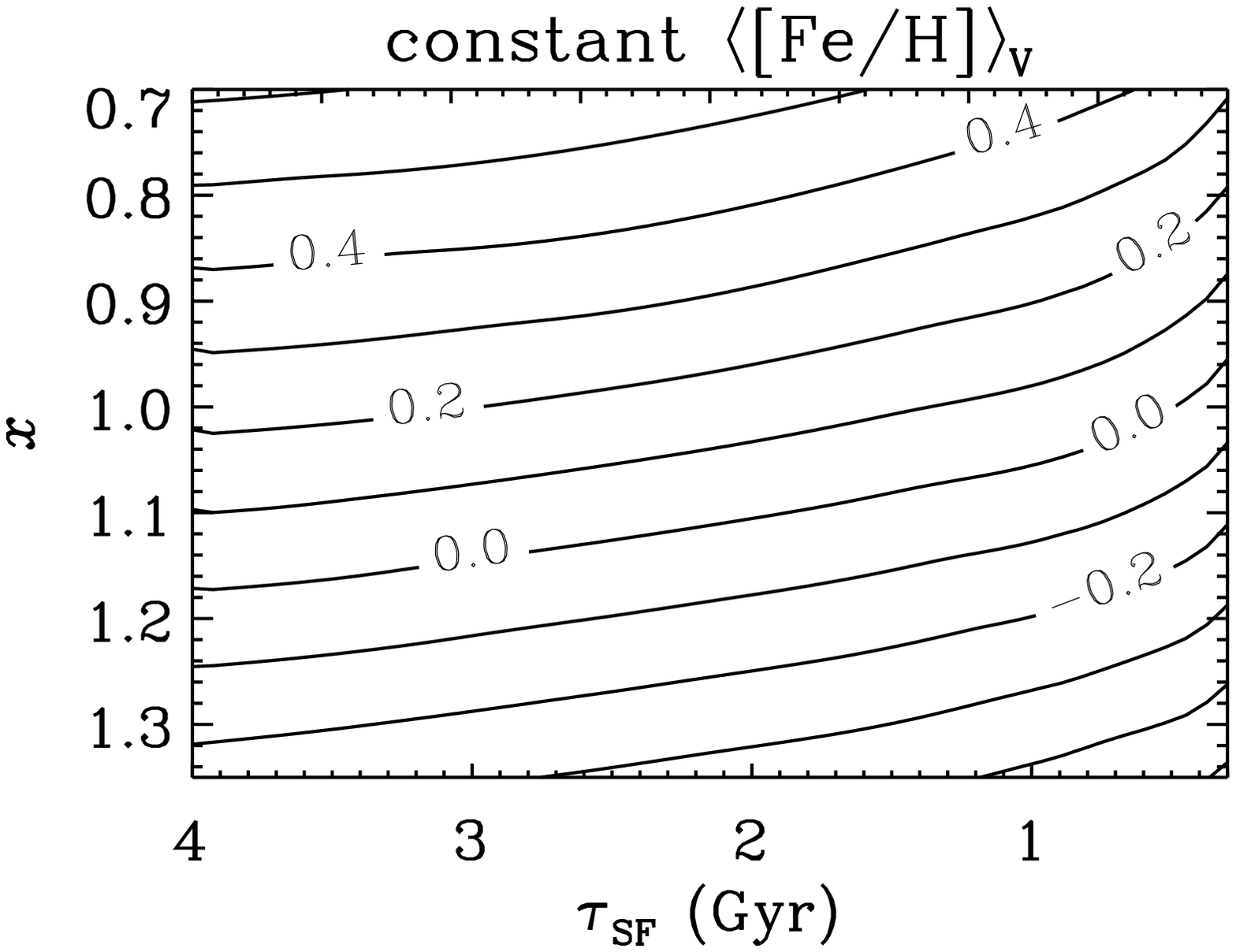,width=\linewidth}
\end{minipage}
\begin{minipage}{0.49\linewidth}
\psfig{figure=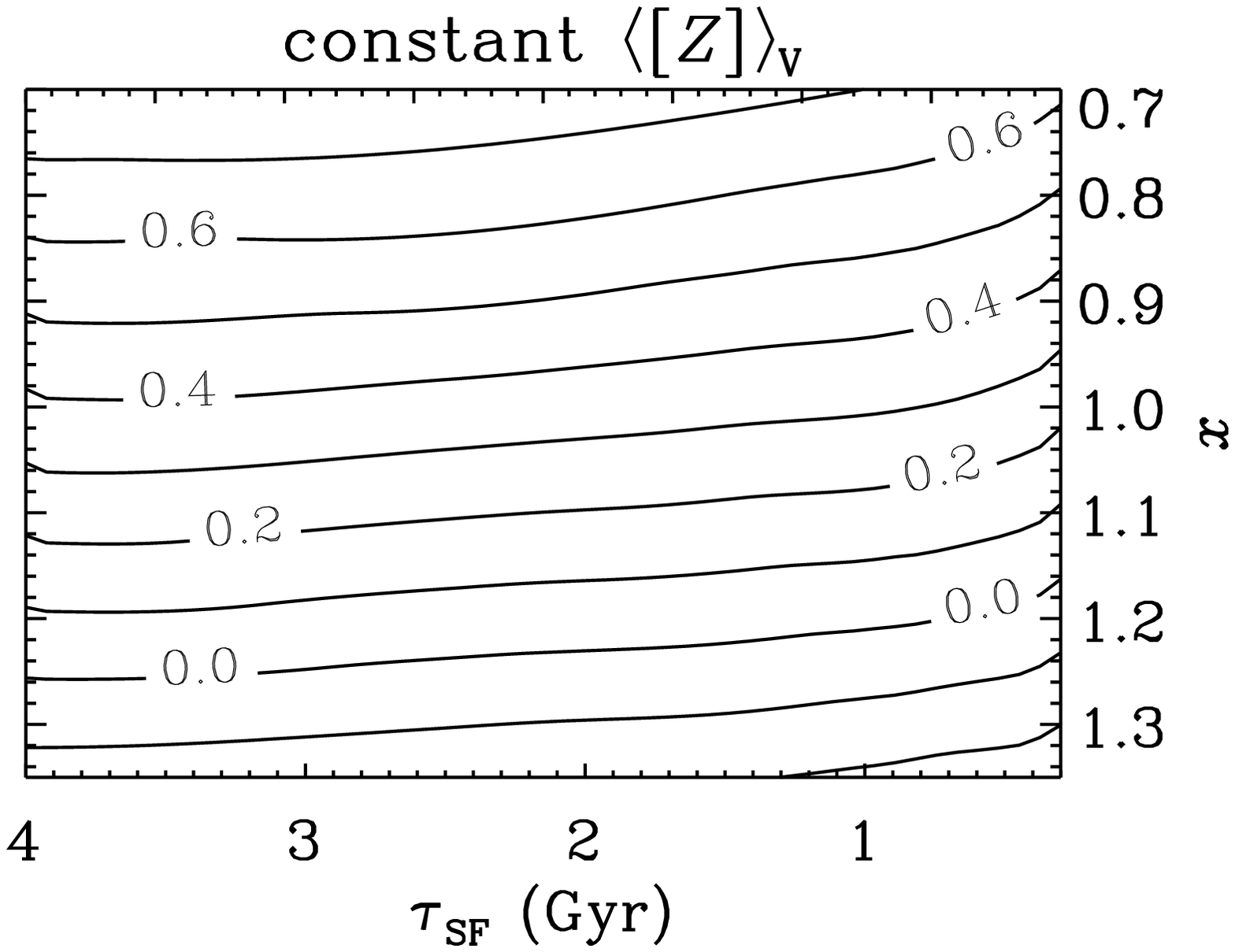,width=\linewidth}
\end{minipage}
\caption[Bla]
{{\em Top-left panel:} The same as Fig.~\ref{fig:ww_surf}, however TNH96
nucleosynthesis is adopted. As a consequence, significantly longer star
formation time-scales ($\la 1$ Gyr) match the observed range of \Ae\
($0.2-0.4$ dex). {\em Top-right panel:} A projection of the contours
indicating constant $\average{[Mg/Fe]}$ to the $\tausf$-$x$-plane. {\em
Bottom panels:} Contours of constant iron abundance (left-hand panel) and
total metallicity (right-hand panel) are plotted as functions of $\tausf$
and $x$. The contours are not independent of $\tausf$ because finite stellar
lifetimes are taken into account (Schaller {\etal\ 1992}). This effect is
larger for iron due to the substantial contribution of SN~Ia to the Fe
enrichment. Generally, the total metallicity is larger than the iron
abundance by $0.1-0.2$ dex. The {\em global} values of both Fe/H and $Z$ turn
out to be {\em sub-solar} ($-0.4$ and $-0.2$ dex, respectively), if an [Mg/Fe]
overabundance is accomplished with Salpeter IMF ($x=1.35$).}
\nocite{SSMM92}
\label{fig:tnh_surf}
\end{figure*}
The results for TNH96 nucleosynthesis are presented in
Fig.~\ref{fig:tnh_surf}. In this case, $\tausf$ as long as a few Gyr still
leads to an [Mg/Fe] overabundance, provided that the IMF is not very steep.
It is important to emphasize that the higher Mg yields in TNH96 basically
stem from a higher contribution by stars in the intermediate mass range
($18-25~\Msun$) which are not severely affected by fall back (Paper~I). The
results are more clearly illustrated in the 2-dimensional panels of
Fig.~\ref{fig:tnh_surf}, where we plot contours of constant
$\average{[Mg/Fe]}$, $\average{[Fe/H]}$, and $\mathaverage{[Z]}$ in the
$\tausf$-$x$-plane. 

The average $Z$ and [Fe/H] are little sensitive to
$\tausf$, while they depend on the IMF slope. This is the mere consequence
of having considered a closed box model which uses up all of the gas. In these
cases the metallicity distributions extend up to the maximum possible value,
which is given by the stellar yields of the SSP, and therefore only depends
on the IMF slope, for a given stellar nucleosynthesis prescription. The
contours are not totally flat because of the relaxation of the instantaneous
recycling approximation which allows a delay in the enrichment. Since a
substantial fraction of Fe comes from SN~Ia the time dependence is larger for
$\average{[Fe/H]}$. The plots also show that the iron abundances are lower
than the actual total metallicity by roughly $0.1-0.2$ dex. Hence at SFTs
of the order one Gyr and below, Fe/H does not trace total metallicity
$Z$ very well. 

From Fig.~\ref{fig:tnh_surf} one can see that $\average{[Fe/H]}\approx 0$ is
achieved with $x$ in the range $1.2-0.9$. For these values of the IMF slope
an [Mg/Fe] overabundance of 0.25 is achieved with $\tausf$ in the range 0.8
to 2.6 Gyr. Restricting to the Salpeter's IMF ($x=1.35$), one needs
$\tausf\approx 0.1-1$ Gyr for an \Ae\ in the range
$\average{[Mg/Fe]}\approx\-0.2-0.3$ dex. The average iron abundance and
metallicity, however, are low reaching $\average{[Fe/H]}\approx -0.4$ and
$\average{[Z]}\approx -0.2$ for $\tausf\approx 1$ Gyr. The fact that the
Salpeter IMF slope predicts sub-solar average metallicities should not be
regarded as demanding flat IMFs for ellipticals. The observational
constraints for super-solar metallicities \cite{Detal87,WFG92} concern the
central parts of the galaxies, while the results discussed here refer to the
global average metallicity of the galaxy. We will concentrate on this
problem in the following subsections.

\smallskip
From now on we restrict our simulations to TNH96 nucleosynthesis,
since, when adopting WW95 yields, the SF time-scales required to accomplish
a [Mg/Fe] overabundance seem unrealistically short.

\subsection{Central abundances for the Clumpy Collapse model}
\label{results_disp}
\begin{figure*}
\psfig{figure=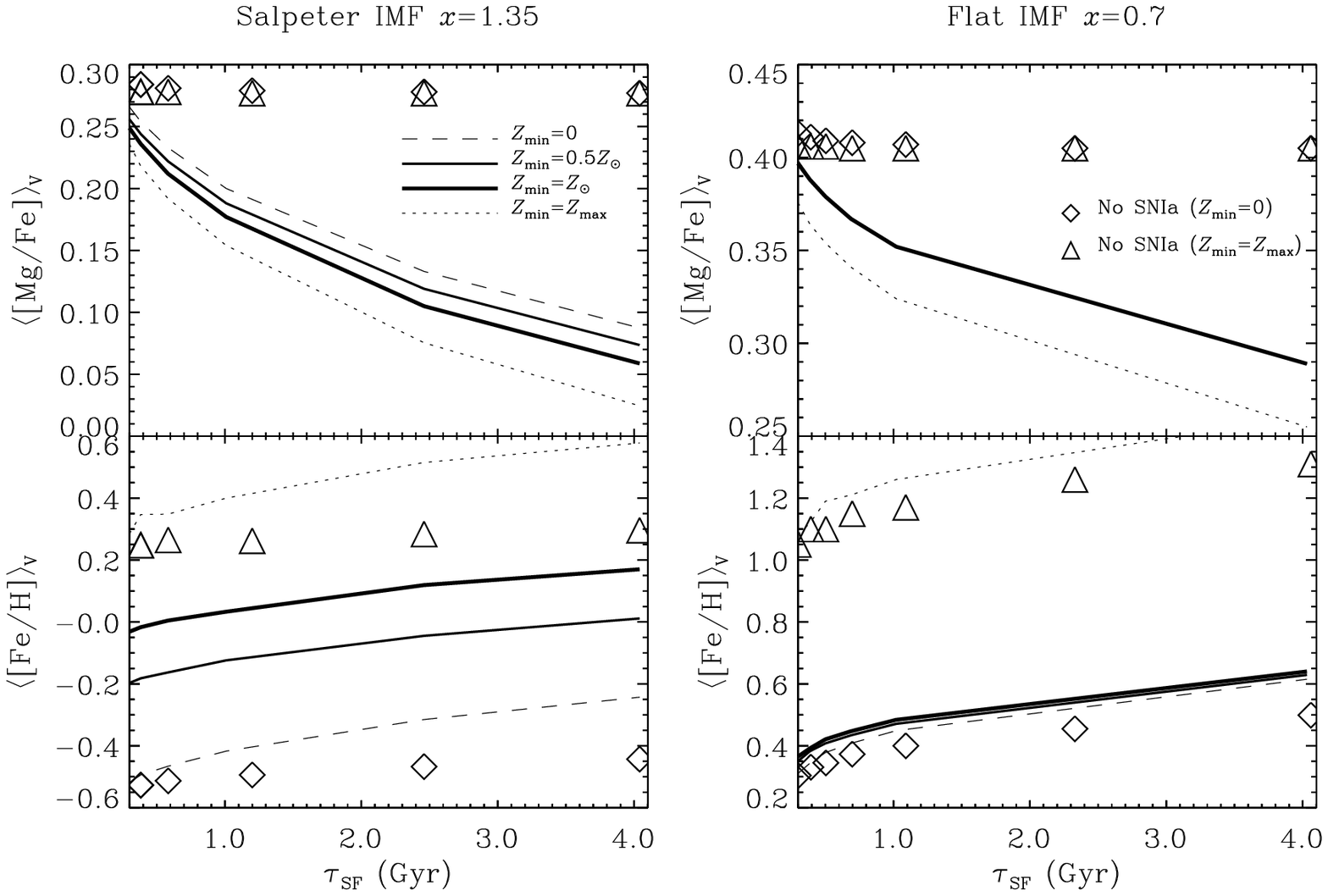,width=\linewidth}
\caption[]
{The figure shows the average stellar [Mg/Fe] overabundance (upper panels)
and iron abundance (lower panels) in the {\em fast clumpy collapse} model as
a function of star formation time-scale for Salpeter IMF (right-hand panel)
and flat IMF (left-hand panel). TNH96 SN~II yields are adopted. The various
line-styles denote different minimum metallicities $Z_{\rm m}$ of the
central stellar population. The symbols represent models excluding Type Ia
supernovae for the two bracketing $Z_{\rm m}=0$ (diamonds) and
$Z_{\rm\-m}=Z_{\rm\-M}$ (triangles). The numbers are additionally shown in
Table~\ref{tab:all} (columns $4-7$).}
\label{fig:disp_class}
\end{figure*}
In this model, the stellar populations dominating the central light of
elliptical galaxies are those in the high metallicity tail of the closed box
distribution. Since the chemical trajectory is characterized by decreasing
overabundance as the metallicity increases, the larger the minimum
metallicity ($Z_{\rm\-m}$) adopted for the CSP inhabiting the central parts,
the lower the average overabundance. This is demonstrated quantitatively in
Fig.~\ref{fig:disp_class}, where we show the predictions for our family of
closed box models with Salpeter IMF (left-hand panel) and flat IMF with
$x=0.7$ (right-hand panel), and varying $\tausf$. The dotted and dashed
lines bracket the possible solutions, showing respectively the results for
$Z_{\rm\-m}=0$ (no segregation of the high $Z$ populations) and
$Z_{\rm\-m}=Z_{\rm\-M}$, which is the highest $Z$ attained by the closed box
model ($Z_{\rm\-M}\sim 3~\Zsun,\ 17~\Zsun$ for $x=1.35$ and $x=0.7$,
respectively). Adopting the Salpeter slope and $\tausf\la 1$~Gyr a
$\average{[Mg/Fe]}\ga 0.2$ dex is obtained for every choice of $Z_{\rm m}$.
The amount of iron locked in stars, instead, is much more sensitive to the
segregation of the populations. In the case of $\tausf=1$~Gyr when cutting
the low-metallicity tail at $0.5~\Zsun$, $\average{[Fe/H]}$ increases from
$-0.44$ to $-0.15$ dex. {\boldmath\bf For the super-solar populations in the
centres of ellipticals, solar iron abundance ($\average{[Fe/H]}=0.01$),
[Mg/Fe] overabundance ($\average{[Mg/Fe]}=0.17$) and super-solar total
metallicity ($\mathaverage{[Z]}=0.18$ dex) are achieved for Salpeter IMF
with a star formation episode lasting 1~Gyr.}

The above example shows that a Salpeter IMF is compatible with the
observations provided that $\tausf$ is shorter than $\sim 1$ Gyr. Longer
$\tausf$ lead to overabundances that seem too small. 

\smallskip
A flatter slope for the IMF obviously relieves the constraint on $\tausf$.
The case of $x=0.7$ is shown in the right-hand panels of
Fig.~\ref{fig:disp_class}. One can see that $\tausf$ as long as 4~Gyr still
yields an overabundance around 0.29 independent of $Z_{\rm m}$. However,
this slope tends also to yield very large [Fe/H] values, which do not seem
implied by the observations. Even with $Z_{\rm\-m}=0$ the average iron
abundance is super-solar. It should be noticed that this is the consequence
of the extremely large $Z_{\rm\-M}$ for this slope, i.e.\
$\approx\-17~\Zsun$. However, there are no known examples of stars with such
high metallicities in any environment.

This `overproduction' of metals can be compensated by transforming a smaller
fraction $f_{\rm trans}$ into stars. Table~\ref{tab:fraction} gives the
results for various $f_{\rm trans}$, and fixed $\tau=1$ Gyr, $x=0.7$,
$Z_{\rm m}=0$.
\begin{table}
\caption{Mean stellar abundances (V-luminosity weighted) for the
{\em fast clumpy collapse} model with $\tausf=1$ Gyr, $x=0.7$ and
$Z_{\rm\-m}=0$, TNH96 yields. The fraction of gaseous mass transformed
into stars is varied from $f_{\rm\-trans}=1.0$ (complete exhaustion) to
$f_{\rm\-trans}=0.5$ to demonstrate the effect on the abundance ratios. The
fifth column gives the total metallicity of the most metal-rich stars
formed.}
\begin{tabular}{ccccc}
\hline
$f_{\rm trans}$ & $\average{[Mg/Fe]}$ & $\average{[Fe/H]}$ &
$\mathaverage{[Z]}$ & $Z_{\rm M}$ \\
1.0 & 0.35 & 0.45 & 0.70 &16.9 \\
0.9 & 0.35 & 0.36 & 0.62 &12.0 \\
0.8 & 0.36 & 0.29 & 0.56 & 9.9 \\
0.7 & 0.36 & 0.19 & 0.48 & 7.9 \\
0.6 & 0.36 & 0.12 & 0.41 & 6.6 \\
0.5 & 0.36 & 0.00 & 0.30 & 5.0 \\
\hline
\end{tabular}
\label{tab:fraction}
\end{table}
For this slope, in order to obtain
$\average{[Fe/H]}\la 0.2$ dex \cite{WFG92,DSP93}, less than
approximately 70 per cent of the total gaseous
mass should be converted into stars, with the rest being lost to the
intergalactic or intracluster medium. Note that the most metal-rich
population exhibits metallicities of $17~\Zsun$ for the case of complete
gas exhaustion.

\subsubsection{Scalo-like IMF}
Assuming an IMF which flattens out for $m\la 0.5~\Msun$
(Scalo 1986; Kroupa, Tout \& Gilmore 1993; Gould, Bahcall \& Flynn 1997)
\nocite{S86,KTG93,GBF97} pushes more mass to massive stars and thus
increases the efficiency of the chemical evolution processing. To
investigate the effect on the results we performed additional simulations
using the estimate of Gould \etal\ at the low-mass end 
\[\begin{array}{ll}
m<0.6~\Msun & \rightarrow\ \phi\sim m^{0.44}\\
0.6~\Msun\leq m\leq 1~\Msun & \rightarrow\ \phi\sim m^{-1.21}
\end{array}\]
and Salpeter slope above $1~\Msun$. The averaged iron abundance -- as
compared to the single slope Salpeter case -- is then further raised to
$\average{[Fe/H]}\approx -0.14$ dex, whereas the Mg/Fe ratio is not
affected. Cutting the metal-rich part of the CSP at $Z_{\rm\-m}=1~\Zsun$ leads
to significantly super-solar abundances in iron
($\average{[Fe/H]}\approx\-0.18$ dex) and total metallicity
($\mathaverage{[Z]}\approx\-0.34$ dex).

If instead a steeper IMF $x=1.7$ above $1~\Msun$ is adopted \cite{S86}, the iron
abundance is slightly increased to $\average{[Fe/H]}=-0.48$ dex. The degree
of \Ae\, however, is reduced to $\average{[Mg/Fe]}=0.1$ dex. A Scalo slope of
$x=1.7$ at the high-mass end requires SFTs of the order $10^8$ yr in order
to produce $\average{[Mg/Fe]}=0.2$ dex. The reason for this pattern lies in
the fact that apart from the contribution of SN~Ia to the Fe enrichment,
only high-mass stars above $8~\Msun$ are responsible for the Mg/Fe ratio.

\subsubsection{The SN~Ia rate}
The time-scales derived up to now are also dependent on the adopted
description of the SN~Ia rate \cite{GR83}. In our models, each stellar
generation gives rise to the first SN~Ia event after 30 Myr from its birth,
and 50$\%$ of the explosions occur within the first 0.5 Gyr approximately.
We take this time as a characteristic time-scale for the Fe enrichment of
the ISM from Type Ia explosions ($\tau_{\rm Fe,Ia}$). The level of the SN~Ia
rate has been calibrated on the chemical evolution of the solar neighborhood
(Paper I). In our modeling, it yields a current rate of $R_{\rm Ia}\approx
0.05-0.1$ SNu\footnote{1SNu= 1SNe$\times (10^{10}L_{\odot})^{-1}\times
(100~{\rm yr})^{-1}$. The luminosities are taken from Worthey
\shortcite{Wo94}.}, depending on the IMF slope and the SFT. This result is in
agreement with the lowest observational estimates of SNe~Ia rates in ellipticals
(see Cappellaro 1996, and references therein)\nocite{C96}.

Given the uncertainties on the SN~Ia progenitors, substantially longer
$\tau_{\rm Fe,Ia}$ could apply, and correspondingly an [Mg/Fe] overabundance
could be accommodated with values of $\tausf$ systematically larger than
derived in the previous section. We notice however that current models
predict values of $\tau_{\rm Fe,Ia}$ not significantly longer than $\sim$ 1
Gyr \cite{Gr96}. It is instructive to evaluate the possibility of
suppressing completely the Fe contribution from SN~Ia, since it also
characterizes the assumption of selective mass loss (see options ii and iv
in the Introduction). The results are shown in Fig.~\ref{fig:disp_class} as
open symbols, where the diamonds refer to the averaged abundances of the
total metallicity distribution, and the triangles to the abundances of the
highest $Z$ stellar population. Note that from stellar abundance
ratios it is not possible to distinguish the options {\em selective mass
loss} and {\em lower SN~Ia rate}. For cluster ellipticals the ICM abundance
in combination with the stellar abundances could be used to distinguish
between the two options, under the assumption that gEs dominantly contribute
to the ICM enrichment.

Without the Fe enrichment from SN~Ia the $\average{[Mg/Fe]}$ saturates at
the maximum SN~II-SSP value of 0.28 dex (Salpeter) and of 0.4 dex ($x=0.7$),
independent of the lower $Z$ cut and the SFT, while $\average{[Fe/H]}$ is
shifted to lower values by roughly 0.2 dex. Total metallicity (not shown in
the figure), instead, is lowered merely by 0.01 dex according to the small
influence of SN~Ia on $Z$ for the SFTs considered here.

Both [Mg/Fe] overabundance and iron abundance become virtually independent
of $\tausf$, since there is no delayed iron enrichment from SN~Ia events.
This computation trivially shows that without the SN~Ia contribution, the
$\alpha$-elements overabundance does not constrain the SFT, while it
constrains the SSP yields, and therefore, having chosen the nucleosynthesis
prescriptions, the IMF slope. Adopting different models for the SN~Ia
progenitors with longer $\tau_{\rm Fe,Ia}$, and/or assuming that a fraction
of the Fe ejected by SN~Ia is not incorporated in the ISM (i.e. selective
mass loss) would lead to intermediate results. Accordingly, longer $\tausf$,
or steeper IMF slopes would meet the observational constraints. A
quantitative exploration of the predictions of different SN~Ia rates, with
the appropriate calibrations on the solar neighborhood data and on the
observed current rates, will be the subject of a forthcoming paper.

\subsection{Central abundances in the model of merging spirals}
\begin{figure*}
\psfig{figure=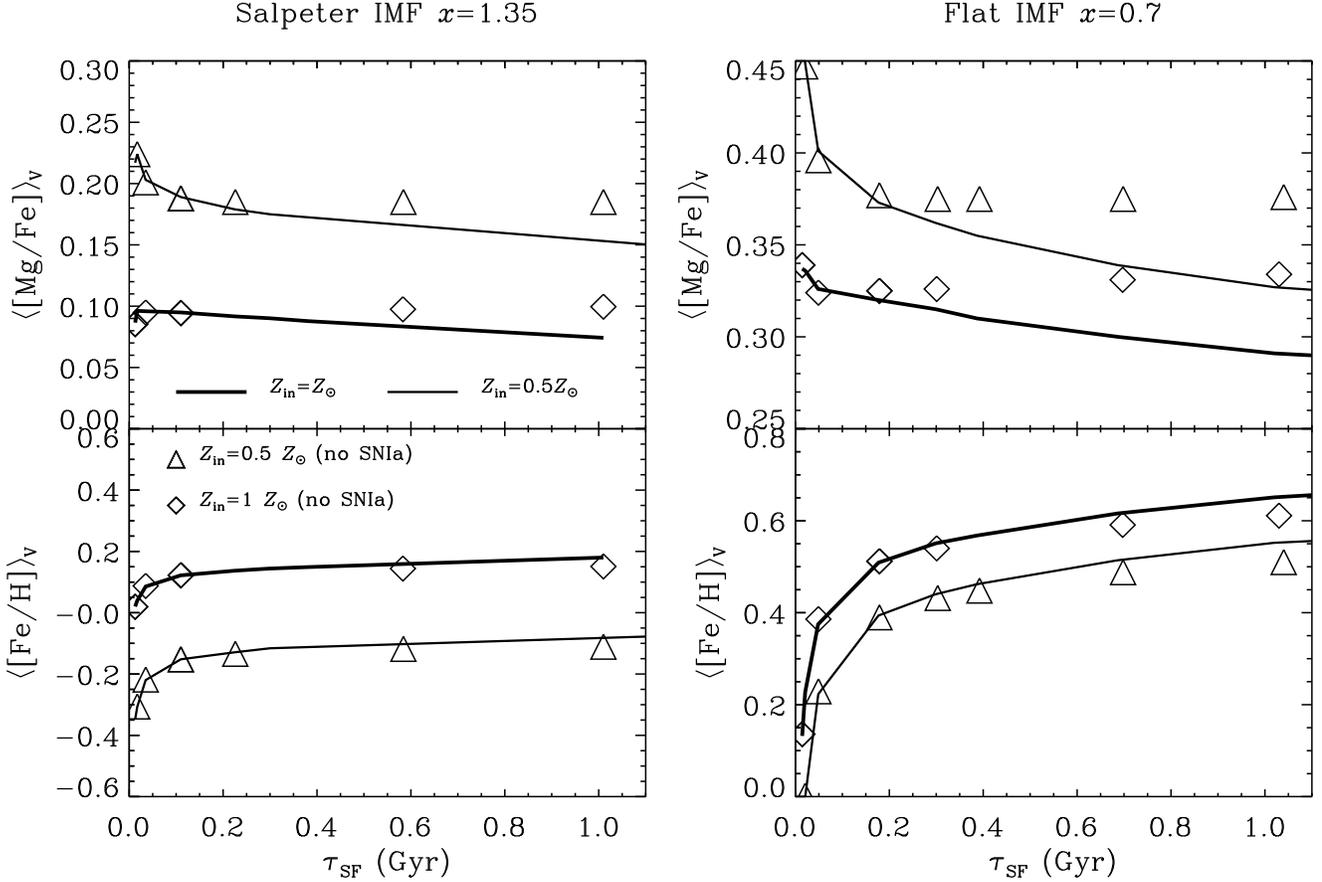,width=\linewidth}
\caption
{The figure shows the average stellar [Mg/Fe] overabundance (upper panels) and
iron abundance (lower panels) in the {\em merging spirals} model as a
function of burst time-scale for Salpeter IMF (right-hand column) and flat
IMF (left-hand column). TNH96 SN~II yields are adopted. The line-styles are
the same as Fig.~\ref{fig:disp_class}, denoting the respective degrees of
pre-enrichment (see also Table~\ref{tab:pres}). The symbols represent models
excluding Type Ia supernovae {\em during the burst} for
$Z_{\rm\-m}=0.5~\Zsun$ (triangles) and $Z_{\rm\-m}=1~\Zsun$ (diamonds). The
numbers are additionally shown in Table~\ref{tab:all} (columns $8-10$).}
\label{fig:disp_merge}
\end{figure*}
As a second possibility for elliptical galaxy formation we consider the
merger of two spirals, and discuss the chemical outcome from a star burst
triggered by the merging event. The main difference with respect to the {\em
fast clumpy collapse} case lies in the initial abundances adopted for the
computation of the chemical evolution. Our model for the solar neighborhood
(Paper~I) predicts that a half solar and a solar metallicity are reached
after 3 and 10 Gyr, respectively. We take the ISM abundances at these ages
as initial conditions for the chemical evolution during the burst (see
Table~\ref{tab:pres}). We recall here that the model in Paper~I adopts
Salpeter IMF, which allows to reproduce the abundance patterns of the solar
neighbourhood.

Modeling the burst of star formation we discuss the burst time-scale
$\tausf$, however the overall SFT of the object includes both $\tausf$ and
the age of the parent spirals (see Fig.~\ref{fig:sfh}). For consistency, we
look at the properties of the final galaxy at a total age of 15 Gyr, i.e.
when the burst populations are 5 Gyr and 12 Gyr old in the cases of solar
and half-solar pre-enrichment, respectively. Note, however, that the passive
evolution of the SPs beyond 1~Gyr only provides a second order effect on the
average abundances. The minimum metallicities of the populations formed at
merging are by construction $Z_{\rm\-m}=0.5~\Zsun$ and $1~\Zsun$, and
represent the stars in the galaxy core. The fractional amount of gaseous
mass is 50 per cent and 20 per cent for the $Z_{\rm\-in}=0.5~\Zsun$ and
$Z_{\rm\-in}=1~\Zsun$ case, respectively.

The results of the calculations for the different parameters considered are
summarized in the last two columns of Table~\ref{tab:all}, and plotted in
Fig.~\ref{fig:disp_merge}, for Salpeter (left-hand panel) and flat IMF
(right-hand panel). Again, the calculations neglecting SNe~Ia are denoted by
the symbols (diamonds: $Z_{\rm\-in}=1~\Zsun$ and triangles:
$Z_{\rm\-in}=0.5~\Zsun$). For Salpeter IMF both models (solar and half-solar
pre-enrichment) fail to produce at the same time high metallicity {\em and}
[Mg/Fe] overabundance. For half-solar initial abundances (characterized by
an initial \Ae\ of 0.1 dex) the resulting average iron abundance is
sub-solar. Starting with solar abundances, instead, super-solar iron
abundance ($\average{[Fe/H]}\approx 0.14$ dex) and metallicity
($\mathaverage{[Z]}\approx 0.20$ dex) are achieved, whereas the [Mg/Fe]
overabundance ($\average{[Mg/Fe]}\la 0.1$ dex) is low. The apparently weak
influence of neglecting SN~Ia (symbols) comes from the fact that in this
model SNe~Ia are excluded only during the short burst phase. The iron
provided by SN~Ia explosions during the evolution of the parent spirals is
included by construction in the initial conditions. Clearly, a flatter IMF
during the burst yields a large overabundance, and high metallicities. It is
worth noticing that the duration of the burst is not important with respect
to the constraints considered here.

\subsection{Comparison between the two models}
We now turn to compare the results in the two extreme modes for the
formation of elliptical galaxies. Inspection of Table~\ref{tab:all} reveals
that, for the same degree of pre-enrichment (i.e. $Z_{\rm\-m}$):
\begin{enumerate}
\item The average metallicity $\mathaverage{[Z]}$ of the
central CSPs is approximately the same for the two models.
\item The $\average{[Fe/H]}$ is systematically larger, and the average
[Mg/Fe] overabundance is systematically lower, in the {\em
merging spirals} model.
\end{enumerate}
The first result simply reflects the fact that in both cases we have an
overall closed box metallicity distribution with the same widths, the
maximum metallicity being determined only by the nucleosynthesis
prescriptions and the IMF slope. The second result is a consequence of the
different initial conditions: in the {\em merging spiral} model all the Fe
ejected by SN~Ia during the past several Gyr of evolution of the parent
spirals is incorporated in the gas undergoing the final burst of SF in the
centre. In the {\em fast clumpy collapse} model, even the high metallicity
stars form early, when pollution by SN~Ia is still relatively modest. The
larger \Ae\ achieved in this model is a direct consequence of the same
effect. Indeed, the differences in both $\average{[Fe/H]}$ and
$\average{[Mg/Fe]}$ are larger for the CSPs with $Z_{\rm m}=1~\Zsun$, with
respect to the cases with $Z_{\rm m}=0.5~\Zsun$. When $\tausf$ is
sufficiently long (e.g. 4 Gyr) the two models converge to similar results
for $\average{[Fe/H]}$ and $\average{[Mg/Fe]}$, since in this case also in
the {\em fast clumpy collapse} model most of the Fe ejected by SN~Ia has
been included in the high metallicity tail of the CSP.


\section{Discussion}
\label{discussion}

\subsection{Fast clumpy collapse vs.\ merging spirals}
The two formation scenarios considered here can be
distinguished by the time-scale on which the pre-enrichment of the central
SPs to solar-like abundances is accomplished.  
In the {\em merging spirals} scenario, this time-scale is as long as
several Gyr, and the
pre-enriched gas that forms new stars has solar element abundances, 
in particular non
$\alpha$-enhanced abundance ratios. 
In the {\em fast clumpy collapse} scenario, instead, the
pre-enriched gas forming the nucleus is overabundant in
$\alpha$-elements because of the preceding short SFT (shorter than 1 Gyr).

Our computations show that for the {\em clumpy collapse} both constraints
for gEs, namely \Ae\ {\em and} super-solar metallicity, can be matched under
the assumption of Salpeter IMF and SFTs up to 1 Gyr. High average
metallicities are accomplished through segregating the metal-rich
populations (with a minimum metallicity around $1~\Zsun$) in the central
region from the rest of the galaxy. 

Super-solar {\em iron} abundance can be obtained if in addition a flattening
of the IMF at the {\em low-mass end} around $0.5~\Msun$ and Salpeter slope
for masses above $1~\Msun$ are applied. Flatter IMFs (in particular at the
high mass end) generally allow longer SFTs, for the same
$\average{[Mg/Fe]}$, and lead to larger average metallicities. The metal
'overproduction' can be avoided by assuming that only a fraction of the
initial gas mass is converted into stars, with consequences on the chemical
enrichment of the intergalactic (or intracluster) medium. Besides, flatter
IMFs, with metallicity distributions extending up to several times $\Zsun$,
do not require the segregation of the high metallicity SPs in the galaxy
cores, to accomplish average metallicities in agreement with the
observational indications. It is worth noticing, however, that there is no
guarantee that these metallicity distributions would yield the strong Mg$_2$
indices observed in the central parts of the galaxies. In fact, the value of
the indices depend on the {\em fitting functions} which seem to saturate at
large metallicity (see G97).

In the case of merging of two Milky Way galaxies, the only way to achieve a
high \Ae\ is to invoke a flat IMF in the final burst. This implies a large
average metallicity for the central SP, with $\mathaverage{[Z]}\approx 0.5$
already for $x$=1. In this case, it seems unlikely that the final burst does
not proceed up to gas exhaustion, since the chemical processing takes place
deep in the potential well. Thus, if ellipticals form in this way, IMF slopes
flatter than $x=1$ seem highly unfavoured from the values of the Mg and Fe
indices observed. The only way out would be to assume very short
time-scales for the final burst of $\taubu\sim 10^7$ yr, which appears
contrived. The merging of two spirals at early times in their chemical
evolution looks more favourable, since our models starting from half solar
initial conditions result in abundances marginally consistent with both
observational constraints (\Ae\ {\em and} high metallicity).

\subsection{Gradients inside the galaxy}
\label{discussion_gradient}
In the {\em fast clumpy collapse} model, in order to accomplish large 
average metallicities in the core population we assumed a cut in the 
$Z$ distribution, such that only stars with $Z > Z_{\rm m}$ are found
in the central parts of the galaxy. Correspondingly the low
metallicity stars (Fig.~\ref{fig:disp_class}) are in the outskirts,
which qualitatively accounts for the
observed metallicity gradients. As a consequence  one expects
that the stars in the outskirts of the elliptical galaxy should be
more $\alpha$-enhanced, with respect to those in the central parts,
i.e. we expect a {\em positive} gradient in the [Mg/Fe] ratio. 
This pattern is present in the solar neighbourhood, with the
metal-poor stars being more $\alpha$-enhanced. 
In the {\rm merging spirals} model, instead, the stars in the outer
parts of the galaxy have approximately solar abundance ratios,
i.e. ellipticals formed in this mode should exhibit {\em negative} gradients
of [Mg/Fe].

Although the derivation of real abundances from line strengths
measurements is highly uncertain (G97; 
Tantalo, Bressan \& Chiosi 1997a, 1997b)\nocite{Gre97,TBC97a,TBC97b},
so far the observations tend to indicate that the \Ae\ is either constant,
or mildly decreasing from the centre to the outer parts of the ellipticals
\cite{DSP93,Me98}. A trend of increasing overabundance with increasing
metallicity (with a constant IMF slope) can only be achieved in a scenario
in which the formation of the metal-rich population is evolutionarily
decoupled from the star formation episode which generated the low
metallicity stars, as assumed in our model of the {\em merging spirals}. In
principle, one could apply this same decoupling also to the {\em fast clumpy
collapse} picture, assuming that the SN~Ia products of the low Z stellar
generation do not mix efficiently with the collapsing gas, and are
preferentially lost to the IGM. However, this looks rather artificial, due
to the very short time-scale over which the formation of the whole galaxy is
accomplished. Anyway, the measurement of abundance gradients offers the
opportunity to discriminate between these two formation scenarios. It is
therefore very important to better understand how the line strengths map
into real abundances, and to study the trend of [Mg/Fe] overabundance with
radius in great detail and in many objects.

\subsection{Caveats and limitations}
\subsubsection{Luminosity vs.\ mass average}
It is usually argued that the difference between luminosity and mass average
is negligible \cite{AY87,M94}. The derived abundances given in this paper
are all weighted by the light emitted in the V-band, the specific
V-luminosities as functions of age and metallicity are adopted from Worthey
\shortcite{Wo94}. In agreement with estimates by Edmunds \shortcite{Ed92},
the metallicity averaged by mass turns out to be systematically higher by
0.1 dex for Salpeter IMF, independent of the SFT:
\[ \Delta[{\rm Fe/H}]\equiv\averagemass{[Fe/H]}-\average{[Fe/H]}\approx 0.1~{\rm dex}\]
Following the argument by Arimoto \& Yoshii \shortcite{AY87} this is due to
metal-poor K giants dominating the light in the V-band (see also Edmunds
1990)\nocite{Ed90}. The effect on the [Mg/Fe] overabundance is significantly
smaller, i.e.:
\[ \Delta[{\rm Mg/Fe}]\equiv\averagemass{[Mg/Fe]}-\average{[Mg/Fe]}\approx -0.01~{\rm dex}\]

A flattening of the IMF lessens the discrepancy between mass and light
average, hence $\Delta[{\rm Fe/H}]\approx 0.05~{\rm dex}$. However Worthey
\shortcite{Wo94} provides models for Salpeter slope only, causing an
inconsistency in our approach. Indeed, at large ages the stellar
$M/L_{\rm\-B}$ ratios are affected by the IMF slope adopted \cite{Ma98}. To
check the influence on our results, we additionally performed simulations
using SSP luminosities derived by Buzzoni \shortcite{B89}, who provides
models also for flatter IMF slope. It turns out that a
$\Delta[{\rm\-Fe/H}]\approx\-0.05~{\rm dex}$ for Salpeter IMF is further
reduced to $\Delta[{\rm Fe/H}]\approx 0.02~{\rm dex}$ with $x=0.7$. We
conclude that within an uncertainty of 0.1 dex for the metallicity the
difference between mass and light average in fact provides a minor effect
independent of the IMF slope.

\subsubsection{Stellar Yields}
Our conclusions are strongly affected by uncertainties of SN~II
nucleosynthesis (Paper~I; Gibson 1997a, 1997b; Portinari, Chiosi \& Bressan
1998)\nocite{TGB98a,G97a,G97b,PCB98}. We analyse the consequence of this
problem by initially considering two different SN~II yields (WW95, TNH96),
but most of the computations are performed with TNH96, since they give the
better fit to the solar neighbourhood data (Paper~I). However, since TNH96
underestimate the Mg/O ratio by roughly 10 per cent (Paper~I), and since the
oxygen yield is likely to be better determined, there is the possibility
that the Mg stellar yield is underestimated in TNH96 models. This leaves
room to further relax the constraints on $\tausf$ and/or to reconcile the
\Ae\ in ellipticals with the local IMF.

In our computations we assume an upper mass cut-off of $40~\Msun$ for both
WW95 and TNH96 SN~II yields. Increasing the upper mass cut-off to
$70~\Msun$, and interpolating between TNH96 yields for a $40~\Msun$ model
and the Nomoto \etal\ \shortcite{Netal97} yields of a $70~\Msun$-star, leads
to an increase of $\sim 0.15$ dex of the [Mg/Fe]-ratio in the SN~II ejecta
in the case of Salpeter IMF (Paper~I). Again, this relieves the constraints
on the short $\tausf$ for the {\em clumpy collapse model}, or on the IMF
slope. In particular, for the case of solar pre-enriched gas (merging Milky
Ways) we performed additional calculations including the $70~
\Msun$
contribution to the ISM enrichment. Indeed, for Salpeter IMF we now obtain
$\average{[Mg/Fe]}\approx 0.2$ dex for $\tausf=0.3$~Gyr . However, the actual
contribution to the ISM from stars more massive than $40~\Msun$ is highly
uncertain (Paper~I), due to the fall-back effect (WW95). Anyway, if these
stars contribute substantially to the ISM pollution, the discrepancy in the
derived overabundance between the {\em fast collapse} and {\em merging
spirals} models becomes larger than what is shown in Table~\ref{tab:all}.
Thus, the conclusion that these two extreme cases of galaxy formation
scenarios lead to different abundance ratios is even reinforced.


\section{Conclusion}
\label{conclusion}
We have explored how to achieve the observed \Ae\ in luminous ellipticals in
the three-dimensional parameter space of star formation time-scale, IMF
slope, and stellar nucleosynthesis. As anticipated in Paper~I on the basis
of SSP yields, we find that, using the stellar yields from WW95, the
observed $\average{[Mg/Fe]}$ are reproduced only with extremely short star
formation time-scales (SFT) of the order $10^7$ yr. Therefore most of the
results presented here have been obtained using TNH96 models.

It turns out that the abundance ratio $\average{[Fe/H]}$ underestimates the
real total metallicity of the system by $0.1-0.3$ dex depending on the
SFT and IMF slope. We also briefly explored the difference between
(V-band) light and mass average metallicity, finding that the mass
averaged [Fe/H] abundance ratios are 
$\approx 0.1$ dex larger than the V-luminosity averaged ones. The
effect on [Mg/Fe] is instead negligible.

\smallskip
We have considered to extreme modes of elliptical galaxy formation, the {\em
fast clumpy collapse} and {\em merging spirals} modes. We find that in the
{\em fast clumpy collapse} model $\average{[Mg/Fe]}\approx 0.2$ dex (in the
high metallicity SPs) is obtained with IMF slopes close to Salpeter,
provided that the overall formation time-scale does not exceed 1 Gyr. For
the {\em merging spirals} model, instead, a sizeable overabundance is
achieved only by claiming a flat IMF for the burst forming the central
population. This is due to the large Fe content of the ISM at the epoch of
the burst. The earlier the merging takes place in the chemical evolutionary
history of the parent spirals, the easier it is to accomplish the
overabundance in the burst population. In general, galaxy formation
time-scales exceeding a few Gyr do not provide $\mathaverage{[\alpha/{\rm
Fe}]}$ ratios consistent with observations, unless a significant deviation
from the local IMF is allowed.

The two extreme models tend to predict opposite trends for the \Ae\
gradient within the galaxy: the {\em fast clumpy collapse} leads
preferentially to larger overabundances in the outskirts where the low $Z$
stars are formed. For the {\em merging spirals} model we expect
approximately solar ratios in the outer stellar populations. If the merging
occurs relatively early in the chemical evolution of the parent spirals, a
slight \Ae\ is expected also in the outer stellar populations. Thus the
behaviour of $\average{[Mg/Fe]}$ with radius can give important hints on the
galaxy formation process. Kauffmann \shortcite{K96} proposed an intrinsic
difference between the formation of cluster and field ellipticals, due to
the high density environment favouring shorter galaxy formation (total)
time-scales. If field objects are formed in a later merging event with
respect to cluster objects we expect that field ellipticals tend to exhibit
lower \Ae\ and steeper \Ae\ gradients than cluster ellipticals.

\section{Acknowledgments}
We would like to thank the referee, M.\ Edmunds, for very interesting
comments on the first version of the paper. We further thank S.~E.\ Woosley
for interesting and enlightening discussions. This work was supported by the
"Sonderforschungsbereich 375-95 f\"ur Astro-Teilchenphysik" of the Deutsche
Forschungsgemeinschaft.



\begin{thebibliography}{}

\bibitem[\protect\citename{Aller \& Greenstein }{1960}]{AG60}
Aller L.~H.,  Greenstein J.~L., 1960, ApJS, 5, 139

\bibitem[\protect\citename{Anders \& Grevesse }{1989}]{AG89}
Anders E.,  Grevesse N., 1989, Geochim.\ Cosmochim.\ Acta, 53, 197

\bibitem[\protect\citename{Arag{\'o}n~Salamanca et~al. }{1993}]{Aetal93}
Arag{\'o}n~Salamanca A., Ellis R.~S., Couch W.~J.,  Carter D., 1993, MNRAS,
  262, 764

\bibitem[\protect\citename{Arimoto \& Yoshii }{1987}]{AY87}
Arimoto N.,  Yoshii Y., 1987, A\&A, 173, 23

\bibitem[\protect\citename{Barnes }{1988}]{B88}
Barnes J.~E., 1988, ApJ, 331, 699

\bibitem[\protect\citename{Barnes \& Hernquist }{1996}]{BH96}
Barnes J.~E.,  Hernquist L., 1996, ApJ, 471, 115

\bibitem[\protect\citename{Bender }{1990}]{Be90}
Bender R., 1990, in Wielen R., ed, Dynamics and Interactions of Galaxies.
\newblock Springer Verlag, Heidelberg, p. 232

\bibitem[\protect\citename{Bender }{1996}]{Be96}
Bender R., 1996, in Bender R.,  Davies R.~L., ed, New Light on Galaxy
  Evolution, IAU Symposium 171.
\newblock Kluwer Academic Publishers, Dordrecht, p. 181

\bibitem[\protect\citename{Bender \& Surma }{1992}]{BS92}
Bender R.,  Surma P., 1992, A\&A, 258, 250

\bibitem[\protect\citename{Bender et~al. }{1992}]{BBF92}
Bender R., Burstein D.,  Faber S.~M., 1992, ApJ, 399, 462

\bibitem[\protect\citename{Bender et~al. }{1996}]{BZB96}
Bender R., Ziegler B.~L.,  Bruzual G., 1996, ApJ, 463, L51

\bibitem[\protect\citename{Buzzoni }{1989}]{B89}
Buzzoni A., 1989, ApJS, 71, 817

\bibitem[\protect\citename{Capellaro }{1996}]{C96}
Capellaro E., 1996, in Bender R.,  Davies R.~L., ed, New Light on Galaxy
  Evolution, IAU Symposium 171.
\newblock Kluwer Academic Publishers, Dordrecht, p.~81

\bibitem[\protect\citename{Chiosi et~al. }{1997}]{Cetal97}
Chiosi C., Bressan A., Portinari L.,  Tantalo R., 1997, A\&A, submitted,
  astro-ph/9708123

\bibitem[\protect\citename{Cole et~al. }{1994}]{Cetal94}
Cole S., Arag{\'o}n~Salamanca A., Frenk C.~S., Navarro J.,  Zepf S., 1994,
  MNRAS, 271, 781

\bibitem[\protect\citename{Colless et~al. }{1998}]{Cetal98}
Colless M., Burstein D., Davies R.~L., McMahan R.~K., Saglia R.~P.,  Wegner G.,
  1998, MNRAS, submitted

\bibitem[\protect\citename{Conti et~al. }{1967}]{Cetal67}
Conti P.~S., Greenstein J.~L., Spinrad H., Wallerstein G.,  Vardya M.~S., 1967,
  ApJ, 148, 105

\bibitem[\protect\citename{Davies }{1996}]{Da96}
Davies R.~L., 1996, in Bender R.,  Davies R.~L., ed, New Light on Galaxy
  Evolution, IAU Symposium 171.
\newblock Kluwer Academic Publishers, Dordrecht, p.~37

\bibitem[\protect\citename{Davies et~al. }{1987}]{Detal87}
Davies R.~L., Burstein D., Dressler A., Faber S.~M., Lynden-Bell D., Terlevich
  R.~J.,  Wegner G., 1987, ApJS, 64, 581

\bibitem[\protect\citename{Davies et~al. }{1993}]{DSP93}
Davies R.~L., Sadler E.~M.,  Peletier R.~F., 1993, MNRAS, 262, 650

\bibitem[\protect\citename{Djorgovski \& Davis }{1987}]{DD87}
Djorgovski S.,  Davis M., 1987, ApJ, 313, 59

\bibitem[\protect\citename{Dressler et~al. }{1987}]{Dretal87}
Dressler A., Lynden-Bell D., Burstein D., Davies R.~L., Faber S.~M., Terlevich
  R.~J.,  Wegner G., 1987, ApJ, 313, 42
  
\bibitem[\protect\citename{Edmunds }{1990}]{Ed90}
Edmunds M.~G., 1990, MNRAS, 246, 678

\bibitem[\protect\citename{Edmunds }{1992}]{Ed92}
Edmunds M.~G., 1992, in Edmunds M.~G.,  Terlevich R., ed, Elements and the
  Cosmos.
\newblock Cambridge University Press, Cambridge, p. 289

\bibitem[\protect\citename{Efstathiou et~al. }{1988}]{Eetal88}
Efstathiou G., Frenk C.~S., White S.~D.~M.,  Davis M., 1988, MNRAS, 235, 7

\bibitem[\protect\citename{Faber et~al. }{1992}]{FWG92}
Faber S.~M., Worthey G.,  Gonz{\'{a}}lez J.~J., 1992, in Barbuy B.,  Renzini
  A., ed, The stellar populations of galaxies, IAU Symposium 149.
\newblock Kluwer Academic Publishers, Dordrecht, p. 255

\bibitem[\protect\citename{Farouki \& Shapiro }{1982}]{FS82}
Farouki R.~T.,  Shapiro S., 1982, ApJ, 259, 103

\bibitem[\protect\citename{Frenk et~al. }{1985}]{Fetal85}
Frenk C.~S., White S.~D.~M., Efstathiou G.,  Davis M., 1985, Nat, 317, 595

\bibitem[\protect\citename{Gehren }{1995}]{G95}
Gehren T., 1995, in Hippelein H., Meisenheimer K.,  R{\"o}ser H.-J., ed,
  Galaxies in the Young Universe.
\newblock Springer, Heidelberg, p. 190

\bibitem[\protect\citename{Gerhard }{1983}]{G83}
Gerhard O.~E., 1983, MNRAS, 202, 1159

\bibitem[\protect\citename{Gibson }{1997a}]{G97a}
Gibson B.~K., 1997a, MNRAS, 290, 471

\bibitem[\protect\citename{Gibson }{1997b}]{G97b}
Gibson B.~K., 1997b, in Cosmic Chemical Evolution, IAU Symposium 187, Kyoto,
  Japan.
\newblock Kluwer Academic Publishers, Dordrecht

\bibitem[\protect\citename{Gibson \& Matteucci }{1997}]{GM97}
Gibson B.~K.,  Matteucci F., 1997, MNRAS, 291, L8

\bibitem[\protect\citename{Gibson et~al. }{1997}]{GLM97}
Gibson B.~K., Loewenstein M.,  Mushotzky R.~F., 1997, MNRAS, 290, 623

\bibitem[\protect\citename{Gorgas et~al. }{1990}]{GAS90}
Gorgas J., Efstathiou G.,  Arag{\'{o}}n~Salamaca A., 1990, MNRAS, 245, 217

\bibitem[\protect\citename{Gould et~al. }{1997}]{GBF97}
Gould A., Bahcall J.~N.,  Flynn C., 1997, ApJ, 482, 913

\bibitem[\protect\citename{Greggio }{1996}]{Gr96}
Greggio L., 1996, in Kunth D., Guiderdoni B., Heydari-Malayeri M.,  Thuan
  T.~X., ed, The interplay between massive star formation, the ISM and galaxy
  evolution.
\newblock Editions Frontieres, Gif-sur-Yvette Cedex-France, p.~89

\bibitem[\protect\citename{Greggio }{1997}]{Gre97}
Greggio L., 1997, MNRAS, 285, 151 (G97)

\bibitem[\protect\citename{Greggio \& Renzini }{1983}]{GR83}
Greggio L.,  Renzini A., 1983, A\&A, 118, 217

\bibitem[\protect\citename{Hernquist }{1993}]{H93}
Hernquist L., 1993, ApJ, 409, 548

\bibitem[\protect\citename{Hernquist \& Barnes }{1991}]{HB91}
Hernquist L.,  Barnes J.~E., 1991, Nat, 354, 210

\bibitem[\protect\citename{Ishimaru \& Arimoto }{1997}]{IA97}
Ishimaru Y.,  Arimoto N., 1997, PASJ, 49, 1

\bibitem[\protect\citename{J{\o}rgensen et~al. }{1995}]{JFK95}
J{\o}rgensen I., Franx M.,  Kjaergaard P., 1995, MNRAS, 276, 1341

\bibitem[\protect\citename{Joseph \& Wright }{1985}]{JW85}
Joseph R.~D.,  Wright G.~S., 1985, MNRAS, 214, 87

\bibitem[\protect\citename{Kauffmann }{1996}]{K96}
Kauffmann G., 1996, MNRAS, 281, 487

\bibitem[\protect\citename{Kauffmann et~al. }{1993}]{KWG93}
Kauffmann G., White S.~D.~M.,  Guiderdoni B., 1993, MNRAS, 264, 201

\bibitem[\protect\citename{Kauffmann et~al. }{1996}]{KCW96}
Kauffmann G., Charlot S.,  White S.~D.~M., 1996, MNRAS, 283, L117

\bibitem[\protect\citename{Kroupa et~al. }{1993}]{KTG93}
Kroupa P., Tout C.~A.,  Gilmore G., 1993, MNRAS, 262, 545

\bibitem[\protect\citename{Larson }{1974}]{L74}
Larson R.~B., 1974, MNRAS, 169, 229

\bibitem[\protect\citename{Lacy et~al. }{1993}]{Letal93}
Lacey C.~G., Guiderdoni B., Rocca-Volmerange B.,  Silk J., 1993, ApJ, 402, 15

\bibitem[\protect\citename{Maraston }{1998}]{Ma98}
Maraston C., 1998, MNRAS, in press

\bibitem[\protect\citename{Matteucci }{1994}]{M94}
Matteucci F., 1994, A\&A, 288, 57

\bibitem[\protect\citename{Matteucci \& Tornamb{\`{e}} }{1987}]{MT87}
Matteucci F.,  Tornamb{\`{e}} A., 1987, A\&A, 185, 51

\bibitem[\protect\citename{McWilliam }{1997}]{McW97}
McWilliam A., 1997, ARA\&A, 35, 503

\bibitem[\protect\citename{McWilliam \& Rich }{1994}]{MW94}
McWilliam A.,  Rich R.~M., 1994, ApJS, 91, 749

\bibitem[\protect\citename{Mehlert }{1998}]{Me98}
Mehlert D., 1998, Phd~thesis, Ludwig-Maximilians Universit{\"a}t, M{\"u}nchen

\bibitem[\protect\citename{Melnick \& Mirabel }{1990}]{MM90}
Melnick J.,  Mirabel I.~F., 1990, A\&A, 231, L19

\bibitem[\protect\citename{Mushotzky et~al. }{1996}]{Metal96}
Mushotzky R., Loewenstein M., Arnaud K.~A., Tamura T., Fukazawa Y., Matsushita
  K., Kikuchi K.,  Hatsukade I., 1996, ApJ, 466, 686

\bibitem[\protect\citename{Negroponte \& White }{1983}]{NW83}
Negroponte J.,  White S.~D.~M., 1983, MNRAS, 205, 1009

\bibitem[\protect\citename{Nomoto et~al. }{1984}]{NTY84}
Nomoto K., Thielemann F.-K.,  Yokoi K., 1984, ApJ, 286, 644

\bibitem[\protect\citename{Nomoto et~al. }{1997}]{Netal97}
Nomoto K., Hashimoto M., Tsujimoto F.-K., Kishimoto N., Kubo Y.,  Nakasato N.,
  1997, Nucl.\ Phys.\ A, 616, 79c

\bibitem[\protect\citename{Peletier }{1989}]{P89}
Peletier R., 1989, Phd~thesis, Rijksuniversiteit Groningen

\bibitem[\protect\citename{Portinari et~al. }{1998}]{PCB98}
Portinari L., Chiosi C.,  Bressan A., 1998, A\&A, 334, 505

\bibitem[\protect\citename{Renzini }{1997}]{R97}
Renzini A., 1997, ApJ, 488, 35

\bibitem[\protect\citename{Renzini \& Ciotti }{1993}]{RC93}
Renzini A.,  Ciotti L., 1993, ApJ, 416, L49

\bibitem[\protect\citename{Renzini \& Voli }{1981}]{RV81}
Renzini A.,  Voli M., 1981, A\&A, 94, 175

\bibitem[\protect\citename{Renzini et~al. }{1993}]{Retal93}
Renzini A., Ciotti L., D'Ercole A.,  Pellegrini S., 1993, ApJ, 419, 52

\bibitem[\protect\citename{Salpeter }{1955}]{S55}
Salpeter E.~E., 1955, ApJ, 121, 161

\bibitem[\protect\citename{Sanders et~al. }{1988}]{Setal88}
Sanders d.~B., Soifer B.~T., Elioas J.~H., Madore B.~F., Matthews K.,
  Neugebauer G.,  Scoville N.~Z., 1988, ApJ, 325, 74

\bibitem[\protect\citename{Scalo }{1986}]{S86}
Scalo J.~M., 1986, Fundam.\ Cosmic Phys., 11, 1

\bibitem[\protect\citename{Schaller et~al. }{1992}]{SSMM92}
Schaller G., Schaerer D., Meynet G.,  Maeder A., 1992, A\&AS, 96, 269

\bibitem[\protect\citename{Soifer \& {et al.} }{1984}]{Setal84}
Soifer B.~T.,  {et al.} , 1984, ApJ, 278, L71

\bibitem[\protect\citename{Tantalo et~al. }{1996}]{Tetal96}
Tantalo R., Chiosi C., Bressan A.,  Fagotto F., 1996, A\&A, 311, 361

\bibitem[\protect\citename{Tantalo et~al. }{1997a}]{TBC97a}
Tantalo R., Bressan A.,  Chiosi C., 1997a, A\&A, submitted, astro-ph/9705060

\bibitem[\protect\citename{Tantalo et~al. }{1997b}]{TBC97b}
Tantalo R., Bressan A.,  Chiosi C., 1997b, A\&A, submitted, astro-ph/9710101

\bibitem[\protect\citename{Thielemann et~al. }{1996}]{TNH96}
Thielemann F.-K., Nomoto K.,  Hashimoto M., 1996, ApJ, 460, 408 (TNH96)

\bibitem[\protect\citename{Thomas et~al. }{1998}]{TGB98a}
Thomas D., Greggio L.,  Bender R., 1998, MNRAS, 296, 119 (Paper~I)

\bibitem[\protect\citename{Truran \& Burkert }{1993}]{TB93}
Truran J.~W.,  Burkert A., 1993, in Hensler G., Theis C.,  Gallagher J., ed,
  Panchromatic View of Galaxies.
\newblock Editions Frontieres, Kiel, p. 389

\bibitem[\protect\citename{van~den Bergh \& Tammann }{1991}]{vBT91}
van~den Bergh S.,  Tammann G.~A., 1991, ARA\&A, 29, 363

\bibitem[\protect\citename{Vazdekis et~al. }{1996}]{Vetal96}
Vazdekis A., Casuso E., Peletier R.~F.,  Beckmann J.~E., 1996, ApJS, 106, 307

\bibitem[\protect\citename{Vazdekis et~al. }{1997}]{Vetal97}
Vazdekis A., Peletier R.~F., Beckmann J.~E.,  Casuso E., 1997, ApJS, 111, 203

\bibitem[\protect\citename{Wallerstein }{1962}]{Wa62}
Wallerstein G., 1962, ApJS, 6, 407

\bibitem[\protect\citename{White \& Rees }{1978}]{WR78}
White S.~D.~M.,  Rees M.~J., 1978, MNRAS, 183, 341

\bibitem[\protect\citename{Woosley }{1986}]{W86}
Woosley S.~E., 1986, in Hauck B.,  Maeder A., ed, Nucleosynthesis and Chemical
  Evolution.
\newblock Geneva Observatory, Geneva

\bibitem[\protect\citename{Woosley \& Weaver }{1995}]{WW95}
Woosley S.~E.,  Weaver T.~A., 1995, ApJS, 101, 181 (WW95)

\bibitem[\protect\citename{Worthey }{1994}]{Wo94}
Worthey G., 1994, ApJS, 95, 107

\bibitem[\protect\citename{Worthey et~al. }{1992}]{WFG92}
Worthey G., Faber S.~M.,  Gonz{\'{a}}lez J.~J., 1992, ApJ, 398, 69

\bibitem[\protect\citename{Ziegler \& Bender }{1997}]{ZB97}
Ziegler B.~L.,  Bender R., 1997, MNRAS, 291, 527

\end{thebibliography}

\end{document}